\documentclass[%
 reprint,
superscriptaddress,
preprintnumbers,
 amsmath,amssymb,
 aps
]{revtex4-2}

\usepackage{graphicx}
\usepackage{dcolumn}
\usepackage{bm}
\usepackage{hyperref}
\usepackage{color}
\usepackage{xcolor}
\usepackage{soul}
\usepackage{slashed}
\usepackage{tikz}
\usepackage{multirow}
\usepackage{cleveref}
\usepackage{comment}
\usepackage{ragged2e}

\hypersetup{
    colorlinks,
    linkcolor={red!60!black},
    citecolor={blue!60!black},
    urlcolor={red!60!black}
}

\newcommand{\be}{\begin{eqnarray}}
\newcommand{\ee}{\end{eqnarray}}
\newcommand{\beq}{\begin{equation}}
\newcommand{\eeq}{\end{equation}}
\def\ltap{\ \raise.3ex\hbox{$<$\kern-.75em\lower1ex\hbox{$\sim$}}\ }
\def\gtap{\ \raise.3ex\hbox{$>$\kern-.75em\lower1ex\hbox{$\sim$}}\ }

\newcommand{\brac}[2]{ \left( \frac{#1}{#2} \right) }

\definecolor{wildstrawberry}{rgb}{1.0, 0.26, 0.64}

\definecolor{puke}{rgb}{0.7, 0.7, 0.4}

\DeclareMathOperator{\Tr}{Tr}
\newcommand{\phicl}{\phi_{c}}

\begin{document}

\preprint{FERMILAB-PUB-25-0100-T, MITP-25-019}

\title{Direct Detection of Ultralight Dark Matter via Charged Lepton Flavor Violation}

\author{Innes Bigaran}
 \affiliation{Department of Physics \& Astronomy, Northwestern University, 2145 Sheridan Road, Evanston, IL 60208, USA}
\affiliation{Theory Division, Fermilab, P.O. Box 500, Batavia, IL 60510, USA}

\author{Patrick J.~Fox}
\affiliation{Theory Division, Fermilab, P.O. Box 500, Batavia, IL 60510, USA}

\author{Yann Gouttenoire}
\affiliation{School of Physics and Astronomy, Tel-Aviv University, Tel-Aviv 69978, Israel}
\affiliation{PRISMA Cluster of Excellence \& Mainz Institute for Theoretical Physics,
Johannes Gutenberg University, Staudingerweg 7, 55099 Mainz, Germany}

\author{Roni~Harnik}
\affiliation{Theory Division, Fermilab, P.O. Box 500, Batavia, IL 60510, USA}

\author{Gordan Krnjaic}
\affiliation{Theory Division, Fermilab, P.O. Box 500, Batavia, IL 60510, USA}
\affiliation{Department of Astronomy and Astrophysics, University of Chicago, Chicago, IL 60637}
\affiliation{Kavli Institute for Cosmological Physics, University of Chicago, Chicago, IL 60637}

\author{Tony Menzo}
\affiliation{Department of Physics, University of Cincinnati, Cincinnati, Ohio 45221, USA}

\author{Jure Zupan}
\affiliation{Department of Physics, University of Cincinnati, Cincinnati, Ohio 45221, USA}

\date{\today}
\begin{abstract} 

We propose a dark matter direct-detection strategy using charged particle decays at accelerator-based experiments. If ultralight $(m_\phi \ll \text{eV})$ dark matter has a misalignment abundance, its local field oscillates in time at a frequency set by its mass. If it also couples to flavor-changing neutral currents, rare exotic decays such as $\mu \to e \phi'$ and $\tau\to e(\mu)\phi'$ inherit this modulation. Focusing on such charged lepton flavor-violating decays, we show that sufficient event samples can enable detection of ultralight dark matter candidates at Mu3e, Belle-II, and FCC-ee.

\end{abstract}

\maketitle

The existence of dark matter (DM) would provide smoking-gun evidence for physics beyond the Standard Model (SM), yet its particle nature remains
unknown despite  decades of dedicated experiments \cite{Cirelli:2024ssz}. While these searches have traditionally focused on
DM masses near the weak scale, in recent years, this global effort has broadened to cover a wide range of DM masses.  Such strategies incorporate techniques from diverse disciplines, including  accelerator \cite{Ilten:2022lfq}, condensed matter \cite{Kahn:2021ttr,Mitridate:2022tnv}, and atomic/molecular/optical physics \cite{Jaeckel:2022kwg}, and their relative advantages depend greatly on the DM mass scale.     

Based on the measured dark matter mass density and the Tremaine-Gunn bound \cite{Tremaine:1979we}, any self consistent theory of ``ultralight" DM
{\it must} be both bosonic and wavelike with a present day classical field value  \cite{PhysRevD.28.1243}
\begin{equation}
\label{eq:phicl:quadratic:case}
\phicl(t)=\phi_0 \cos(m_\phi t +\delta),
\end{equation}
where $m_\phi$ is the DM mass, $\delta$ is an arbitrary phase. 
The amplitude satisfies  
\beq\label{eq:phi0DM}
\phi_0=\frac{\sqrt{2\rho_{\phi}}}{m_\phi}\simeq 2.5\,\text{TeV} \biggr(\frac{10^{-15}\,\text{eV}}{m_\phi}\biggr)~,~~
\eeq
where $\rho_\phi = 0.4$ GeV/cm$^{3}$  is the local dark matter mass density \cite{deSalas:2020hbh} and $\phi$ is said to be ``misaligned" from its potential minimum~\cite{Preskill:1982cy,Abbott:1982af,Dine:1982ah}. 
 This unique\footnote{
 If the functional form were to substantially deviate from Eq.~\eqref{eq:phicl:quadratic:case}, the energy density would not redshift like non-relativistic pressureless matter \cite{PhysRevD.28.1243}, which would spoil cosmological structure formation.  Alternatively, if $\phi$ were a spin-1 field, Eq. \eqref{eq:phicl:quadratic:case} would feature a polarization vector, but the time dependence and the amplitude would not be affected. } description of ultralight DM offers a powerful first-principles starting point for {\it any} investigation into DM candidate masses in the range $10^{-21}\,{\rm eV} < m_\phi < 1\, {\rm eV}$.  Masses below $10^{-21}\,{\rm eV}$ are excluded  because the DM de-Broglie wavelength exceeds the scale of dwarf galaxies~\cite{Zimmermann:2024xvd,Benito:2025xuh}.

As the field in Eq.~\eqref{eq:phicl:quadratic:case}  modulates in time with frequency $m_\phi$, interactions between $\phi$ and the SM fields can inherit the characteristic periodicity
\be
\label{eq:tau_phi}
\tau_\phi = \frac{2\pi}{m_\phi} \simeq 4 \, {\rm s} \brac{10^{-15} \rm eV}{m_\phi},
\ee
which may be appreciable on experimental timescales. 
While there is a large literature on DM-induced time variation in fundamental constants, particle masses,  and nuclear spin-precession, these strategies assume flavor blind interactions between $\phi$ and SM fields (see \cite{Antypas:2022asj,Hui:2021tkt,Safronova:2017xyt} for reviews). Ultralight DM has been studied as a catalyst for time variation in the quark and lepton  mixing matrices \cite{Berlin:2016woy,Brdar:2017kbt,Krnjaic:2017zlz,Capozzi:2018bps,Dev:2022bae,Losada:2023zap,Dine:2024bxv}, in the CP phase of strong interactions~\cite{Lee:2020tmi,Zhang:2023lem,Alda:2024xxa}, and of fundamental constants (see e.g.~\cite{Bouley:2022eer}). 
However, to date, no study has explored its possible impact on flavor violating processes. 

In this {\it Letter}, we introduce a novel DM search strategy based on 
flavor changing neutral current (FCNC) transitions induced by the presence of a misaligned field. We focus on charged lepton flavor violating (CLFV) decays, and find that certain classes of dark-visible interactions can viably induce time-modulation at levels that can be probed at existing and future experiments. 

\begin{figure*}[t!]
\centering
\hspace{-0.5cm}
\includegraphics[width=0.5\textwidth]{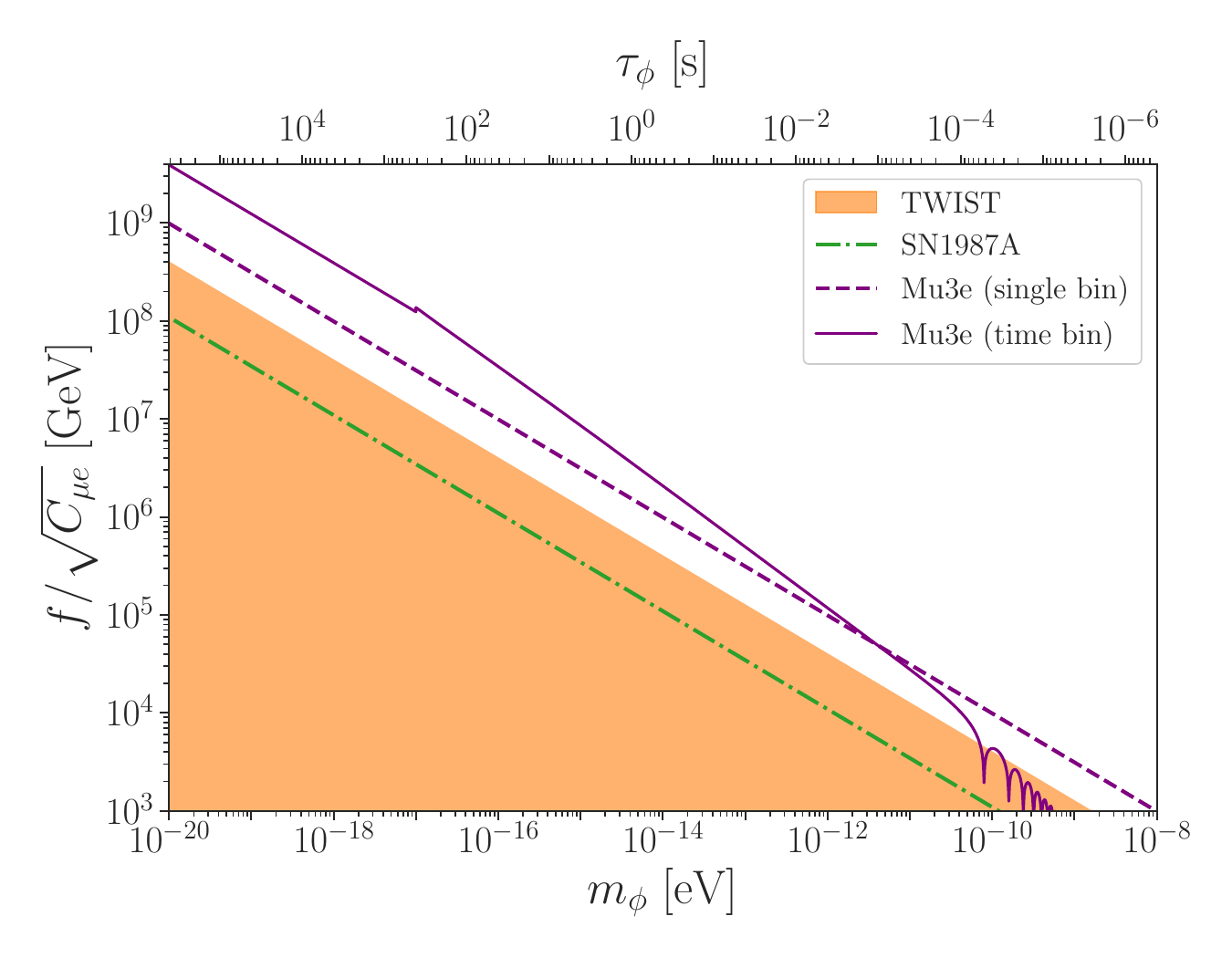}~~
\includegraphics[width=0.5\textwidth]{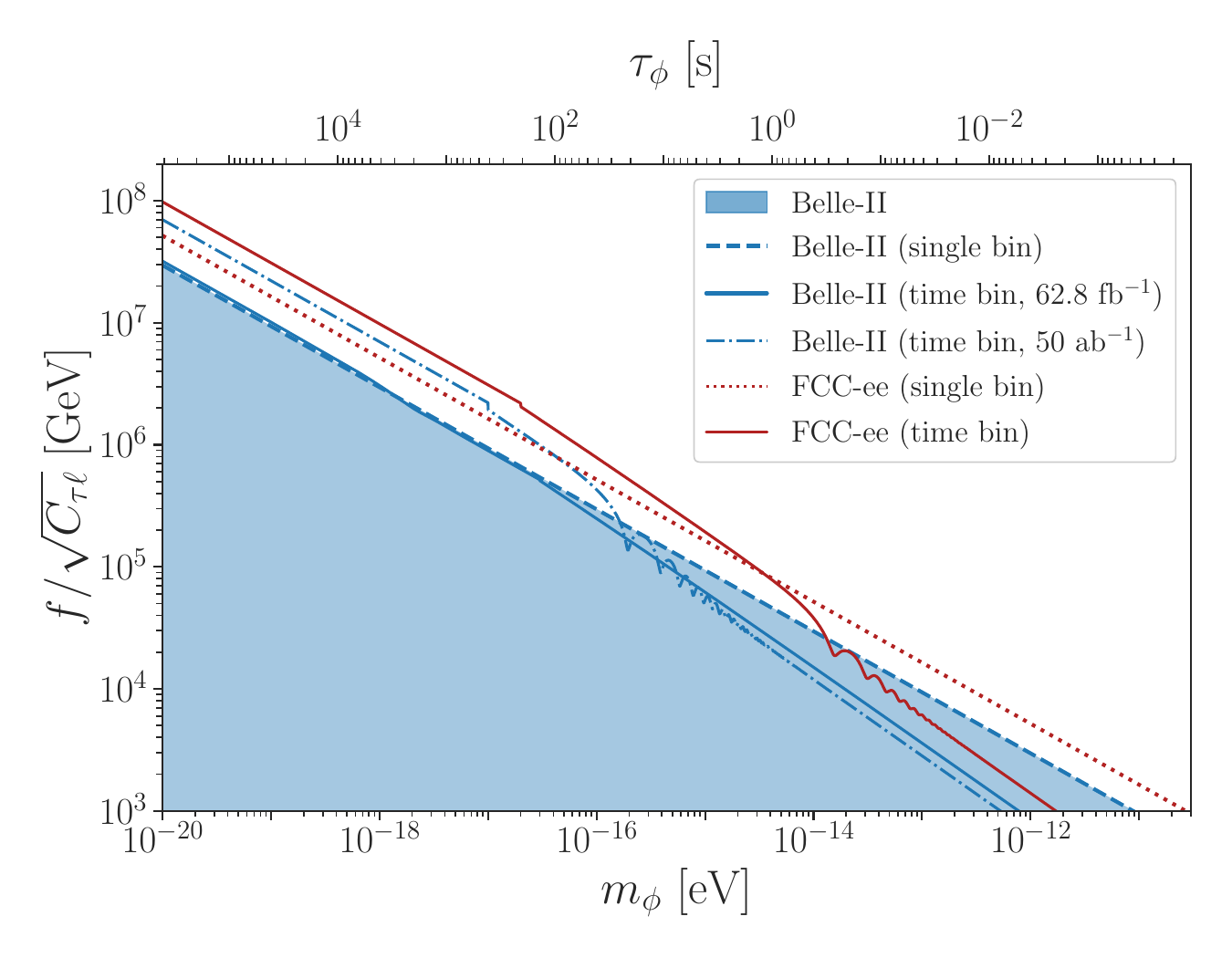}
\caption{
Present and projected constraints on LFV couplings $f/\sqrt{C_{\mu e}}$ (left) and $f/\sqrt{C_{\tau \ell}}$ (right) for time-binned and single-bin analyses, as a function of the scalar mass, $m_\phi$ -- see Eqs.~\eqref{eq:Lint:main} and \eqref{eq:li:to:lj:phi}. \textbf{Left}: Present constraint from TWIST~\cite{TWIST:2014ymv,TWIST:2004hce} is shaded in yellow. The projected single-bin (dashed line) and time-binned (solid line) analysis reach from Mu3e~\cite{Perrevoort:2018okj, Perrevoort:2023qhn} are shown in purple. The astrophysical bound on $f/\sqrt{C_{\mu e}}$ from SN1987A~\cite{Raffelt:1990yz,Calibbi:2020jvd} is shown here as a dot-dashed green line -- see main text for details. \textbf{Right}: The constraints on $f/\sqrt{C_{\tau \ell}}$ from present and projected Belle-II analyses~\cite{Belle-II:2022heu} are shown here in blue: current 62.8 fb$^{-1}$ time-integrated constraint is shaded, single-bin constraint is a dashed line and projected time-binned constraint is a solid line; the 50 $\text{ab}^{-1}$ projected time-binned constraint is a dot-dashed line. The projected constraints from FCC-ee Tera-Z run \cite{Bernardi:2022hny} are shown in red: dotted line for single-bin analysis, and red solid line for time-binned analysis.
The constraints on $C_{\tau \mu}$ and on $C_{\tau e}$ (denoted collectively here as $C_{\tau \ell}$) almost completely overlap on this logarithmic scale.
}
\label{fig:test}
\end{figure*}

As numerical examples, we show projections for a possible reanalysis of existing Belle-II data \cite{Belle-II:2018jsg,Belle-II:2022heu}, as well as the estimates for future   Mu3e~\cite{Mu3e:2020gyw,Perrevoort:2018okj, Perrevoort:2023qhn} and  FCC-ee  \cite{FCC:2018byv,Bernardi:2022hny,Dam:2018rfz} sensitivities -- these results are presented in Fig.~\ref{fig:test}. Similar analyses could be performed at other current and future CLFV experiments,  including MEG-II \cite{MEGII:2018kmf,Calibbi:2020jvd,Jho:2022snj}, Mu2e \cite{Mu2e:2008sio,Mu2e:2014fns,Bernstein:2019fyh,Hill:2023dym,Knapen:2023zgi}, COMET \cite{COMET:2018auw,Kuno:2013mha,Hill:2023dym}, BES-III \cite{BESIII:2020nme,BESIII:2009fln}, Super Tau Charm Factory \cite{Pich:2024qob,Achasov:2024eua,Achasov:2023gey}, and CEPC \cite{Ai:2024nmn}.

\medskip

\noindent{\bf Time-dependent CLFV interactions.} 
Arguably the simplest interactions of $\phi$ to the SM fermions arises from the dimension-5 Higgs interaction $\phi H \bar L_i \ell_j$.  Once the Higgs acquires its vacuum expectation value this operator becomes $y_{ij} \phi \, \overline \ell_i \ell_j$ and leads to decays of the charged lepton to dark matter.  However, this interaction does \emph{not} lead to observable time-dependent CLFV transitions, due to several challenges that we outline below.  \begin{enumerate}

\item  {\em Time-independent Limits:}
The off-diagonal interactions in the dimension-5 operators $\phi H \bar L_i \ell_j$  predict flavor changing processes, including the decays: $
\tau\to \mu\phi, \tau \to  e\phi $, and  $\mu\to e\phi $,
where the final state $\phi$ is invisible to experiments. For such linear $\phi$ couplings the decay rates are \emph{independent} of the $\phi_c$ background at leading order and, therefore, do not inherit time modulation. 
Furthermore, conventional time-averaged searches for $\ell_i \to \ell_j + $ invisible decays yield stringent limits on the couplings~\cite{Belle-II:2022heu,Calibbi:2020jvd,Jodidio:1986mz, TWIST:2014ymv}
\be
\label{eq:ybounds}
~~~~~~~~~|y_{\tau e}|, |y_{\tau \mu}| \lesssim   2 \times  10^{-7}~,~ |y_{\mu e}| \lesssim 10^{-10},
\ee
which severely limit any CLFV time-dependent effects  that  could arise at higher orders, via additional $y_{ij}$ insertions. 

 \item {\em Mass-Shift Effects:}
In the case of the Higgs operator, $\phi H \bar L_i \ell_j$, setting $\phi$ to its background value $\phi_c$  yields corrections to  the lepton mass matrix:
\be
\label{eq:mass-time}
~~~~~~~~~~m_{ij}(t)={\rm diag}(m_e, m_\mu, m_\tau)+y_{ij}\phi_c(t)~,
\ee 
where the additional mass terms induce time dependence in SM decays. For example, the branching ratio ${\cal B}(\mu \to e \nu \bar \nu)$ acquires time modulation  with an amplitude $\propto {\mathcal O}(y_{\mu\mu}, y_{\mu e}^2)$, due to the modified muon mass.  Furthermore, these terms can also induce time-dependent exotic FCNC decays, such as $\mu\to e\gamma$ or $\tau \to e\gamma$. However, the  branching fractions for these processes are suppressed by both  $y_{ij}^2$ and by
$\dot \phi_c$.\footnote{In the $\tau_\phi \gg \tau_\ell$ limit the $\phi_c$ term in Eq.~\eqref{eq:mass-time} merely corrects the charged lepton mass matrix, and does not result in a new source of flavor violation.} Thus, after taking into account the constraints from Eq.~\eqref{eq:ybounds}, all time-varying CLFV signals are unobservably small. 

\end{enumerate}
\noindent Similarly, a dimension-5 axion portal operator of the form $\partial_\mu \phi \bar{\ell}_i \gamma^\mu \gamma_5 \ell_j$ leads to time-independent dark matter decays of charged leptons.  

For the charged lepton decays to be time dependent the operator must involve two insertions of dark sector fields, with at least one of these insertions being the ultra-light dark matter.  If both of the insertions are the dark matter, operators of the form $\phi^2 \bar \ell_i \ell_j$ 
then the decays can become time dependent as the decay amplitude is proportional to $\phi_c$.  However, this raises the problem that loop corrections to $m_\phi$ depend explicitly on the local DM density, $\phi_c$, leading to \emph{position dependent} fine tuning.  This problem does not exist if there are two separate fields only one of which has a classical field value.

As a concrete example we will consider interactions of the form
\beq
\label{eq:Lint:main}
{\cal L}_{\rm int}\supset    \frac{i \phi \,(\partial_\mu \phi')}{2 f^2}    \,  \overline \ell_i \gamma^\mu ({C}^{V}_{ij} + {C}^{A}_{ij} \gamma_5 ) \ell_j +{\rm h.c.},
\eeq
where $\phi$ is DM and $\phi'$ is another ultra-light dark sector field,
 $C^{V,A}_{ij}$ are hermitian matrices in flavor space and the $i,j$ indices run over all charged fermion species and $f$ is the effective energy scale associated with the generation of this operator.  In the Supplemental Material we show that the interaction in Eq.~\eqref{eq:Lint:main} can be the leading DM-SM interaction if $\phi$ and $\phi'$ are pseudo-Nambu-Goldstone bosons of a spontaneously broken non-Abelian global symmetry, and how $f$ relates to the physics that generates this operator.
In the Supplemental Material~\cite{SuppMat} we also give further examples of such quadratic interactions with the dark sector.

\begin{figure}
    \centering
\includegraphics[width=0.5\linewidth]{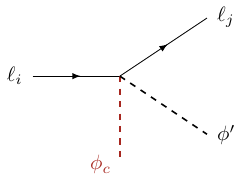}
    \caption{Model contribution to $\ell_i\to \ell_j \phi'$, where the interaction with the background classical field ($\phi_c$) is denoted in red.}
    \label{fig:FM}
\end{figure}

Since the interaction in Eq.~\eqref{eq:Lint:main}  contains two light field insertions, expanding $\phi$ around its background value, $\phi_c$, results in time-dependent FCNC decays at leading order. To make this time modulation manifest, we integrate by parts and replace the $\phi$ field with $\phi_c$ to obtain 
\be
\label{eq:Lint:main2}
{\cal L}_{\rm int}\! \supset \!
 \frac{  \phi_{c}  }{2 f^2} 
\, \phi' \, \overline \ell_i
 \Big[
{ C}_{ij}^V (m_{\ell_i} \!- \! m_{\ell_j} ) 
+{ C}^A_{ij} (m_{\ell_i} \!+\! m_{\ell_j} )       \gamma_5  \Big]\ell_j
,\,~~~
\ee
 where $m_{\ell_i}$ is the mass of $\ell_i$. 
 The branching fraction for $\ell_i \to \ell_j \phi'$ decays is now explicitly calculated from the diagram in Fig.~\ref{fig:FM} and is time-dependent,
  \be
\label{eq:li:to:lj:phi}
{\cal B}(\ell_i\to \ell_j\phi') = \frac{C_{ij}^2 \phi_0^2}{64\pi  f^4} 
\frac{m_{\ell_i}^3 }{\Gamma_{\ell_i}}
 \cos^2(m_\phi t +\delta),
\ee
where $C_{ij}^2\equiv
\big|C^{V}_{ij}\big|^2 +
\big|C^{A}_{ij}\big|^2 $,
$\Gamma_{\ell_i}$ is the total width of lepton $\ell_i$,
$\phi_0$ is the field amplitude from Eq.~\eqref{eq:phi0DM}, 
and we have approximated all final state particles as massless except for the appearance of $m_\phi$ in the cosine.  Note that, as part of the dark sector, $\phi'$ 
will be invisible on accelerator length scales.

\medskip

\noindent{\bf Observing periodic signals.}
New physics transitions $\ell_i\to \ell_j \phi'$ have an irreducible SM background from the tree-level $W$-boson mediated decays, $\ell_i \to \ell_j \nu_i \bar \nu_j$. Due to three-body kinematics, this background mimics the signal $\ell_i\to \ell_j \phi'$ \emph{only} when the invariant mass of the two neutrinos is close to zero, within experimental resolution (the $\phi'$ mass is assumed to be far below the invisible mass resolution in these experiments). Time-dependent signals offer a valuable handle to distinguish signal from background, particularly because the background events do not exhibit periodic modulation. Observing a time modulation in $\ell_i\to \ell_j \phi'$ decays would constitute compelling evidence for DM or another misaligned field with nontrivial cosmological abundance.\footnote{In general, 
$\phi$ could be a subdominant DM component, and still lead to a time-dependent signal.}.There is a non-time-dependent decay of $\ell_i \to \ell_j \phi \phi'$ which is hard to separate from background since the energy of the final state lepton closely mimics the SM decay.

\begin{figure*}[t!]
\includegraphics[width=0.47\textwidth]{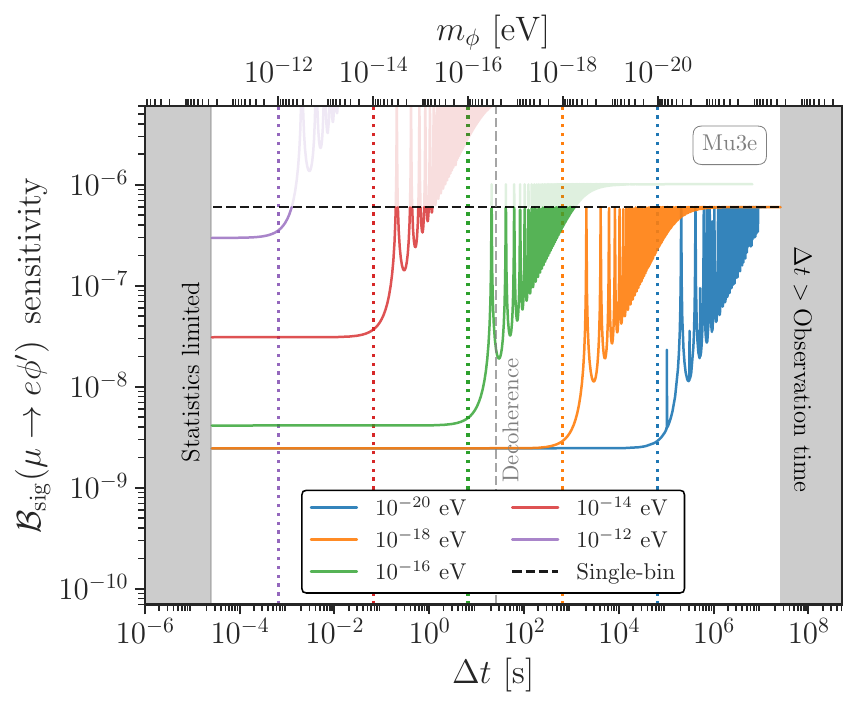}\hspace{0.1in}
\includegraphics[width=0.47\textwidth]{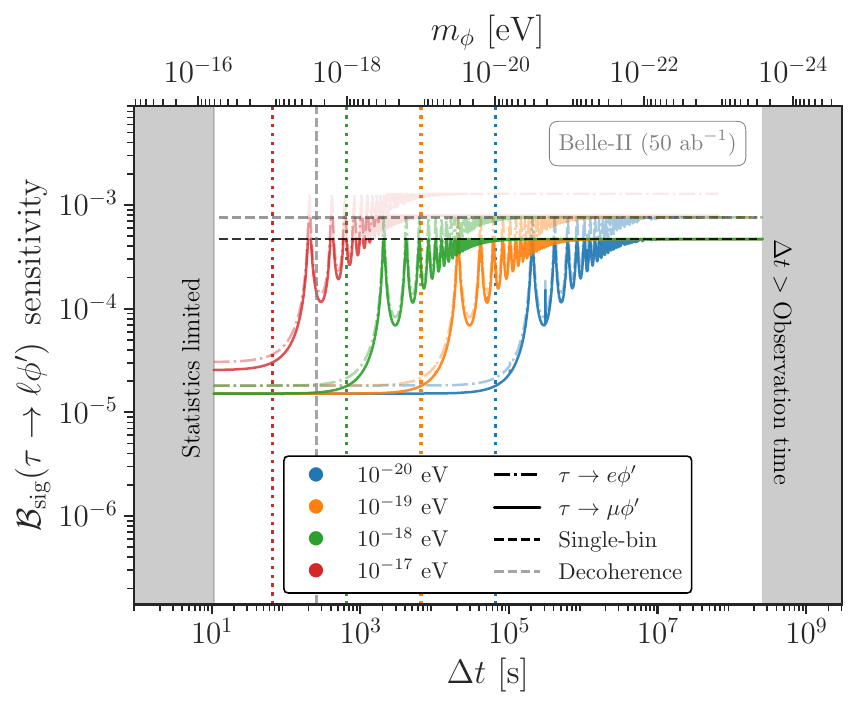}
  \caption{The 90\% confidence upper limit on the signal branching $\mathcal{B}_{\rm sig}$ obtained from a time-dependent analysis,  as a function of bin size $\Delta t$ (solid, dashed-dotted lines) for different $\phi$ masses at Mu3e (left) and Belle II {using the full projected integrated luminosity} (right), using the same assumptions as in Fig. \ref{fig:test}. 
  The dotted vertical lines denote approximately when the oscillatory signal for each mass is resolved, $\Delta t \sim m_\phi^{-1}$. The horizontal dashed black line denotes the upper limit for a time-independent, single-bin, analysis. 
  }
\label{fig:upper_limit_vs_delta_t}
\end{figure*}

{There are many possible analysis techniques to determine the statistical sensitivity to the sinusoidal signals in Eq.~\eqref{eq:li:to:lj:phi} in the presence of substantial SM backgrounds.  We follow a simplified approach, using the  binned $\chi^2$ when possible and the unbinned Rayleigh periodogram elsewhere.\footnote{We thank an anonymous referee for advocating we extend the binned analysis.}  Furthermore, we assume } continuous data collection over total observation time $T$.  For the $\chi^2$ analysis, the data are divided into $n_{\rm bin}=T/\Delta t$ equal-sized bins, each of duration $\Delta t$, which are longer or equal to the experimental resolution.
Using Eq.~\eqref{eq:li:to:lj:phi}, the predicted event rate is 
\begin{equation}
\label{eq:Ndot}
\dot{N}_{\rm pred}(t) = \frac{N_{\rm tot}}{T} \big[\mathcal{B}_{\rm bg} f_{\rm bg} 
  + 2 \mathcal{B}_{\rm sig} f_{\rm sig} \cos^2 (m_\phi t + \delta) \big],
\end{equation}
where $N_{\rm tot}$ is the total number of $\ell_i$ decays, $\mathcal{B}_{\rm bg(sig)}$ and $f_{\rm bg(sig)}$ are the  background (signal) branching fractions and experimental efficiencies, respectively. Local DM with velocity $v \sim 10^{-3}$ sets a characteristic coherence time
\be
\tau_{\rm coh} \sim \frac{1}{m_\phi v^2} \simeq
 7 \times 10^5 \, {\rm s} \brac{10^{-15} \rm eV}{m_\phi} \brac{10^{-3}}{v}^2 \!,~~
\ee
during which the classical DM field maintains the same phase $\delta$. For $T > \tau_{\rm coh}$, an experiment can experience two or more coherent `patches' with different phases. Further details about decoherence in our analysis may be found in the Supplemental Material~\cite{SuppMat}.
In Eq.~\eqref{eq:Ndot}, the \textit{time-averaged} branching ratio for the signal  channel satisfies
 \be
 \mathcal{B}_{\rm sig} 
 \equiv \langle{\cal B}(\ell_i\to \ell_j\phi') \rangle
 = \frac{ 
 C_{ij}^2  }{128\pi   } \frac{\phi_0^2 }{f^4}\frac{ m_{\ell_i}^3 }{\Gamma_{\ell_i}}~,
 \ee
where we have used Eq.~\eqref{eq:li:to:lj:phi}.
 The number of background events, assumed to be non-modulating and dominant over the signal, is given by
\be
N_{\rm bg}=N_{\rm tot} \mathcal{B}_{\rm bg} f_{\rm bg}~~~,~~~\mathcal{B}_{\rm bg} f_{\rm bg} \gg \mathcal{B}_{\rm sig} f_{\rm sig}.
\ee
The number of events per time bin is obtained from integrating $\dot{N}_{\rm pred}(t)$ over each bin interval $\Delta t$. Using Asimov datasets\footnote{In an Asimov data set, all observed quantities are set
equal to their expected values. } we  construct a $\chi^2$ test statistic, which depends on the quadratic sum of statistical and systematic errors in each bin, which  respectively satisfy
\be
\label{eq:sigmas}
 \sigma_{\rm stat} = \sqrt{N_{\rm bg}/n_{\rm bin}}~~,~~ \sigma_{\rm sys} = \alpha N_{\rm bg}/n_{\rm bin},
\ee
where $\alpha$ is the relative systematic uncertainty. The $\chi^2$ test statistic can then be translated to the expected bounds on $f/\sqrt{C_{ij}}$ from Eq.~\eqref{eq:li:to:lj:phi} using time-dependent $\ell_i\to \ell_j \phi'$ searches.  
As shown in the Supplemental Material~\cite{SuppMat},
in the ``fine-binned" (fb) limit where $m_{\phi}\Delta t \ll 1$, $m_\phi T \gg 1$, the time-dependent test statistic $\chi^2_{\rm fb}$ can be written 
\be\label{eq:finebinning-main}
\chi^2_{\rm fb} =  \frac{3}{2} \left[ 1+
\frac{n_{\rm bin}}{3}
\left(\frac{\sigma_{\rm sys}}{\sigma_{\rm stat}}  \right)^2\right]\chi^2_{\rm const}~,
\ee
where $\chi^2_{\rm const}$ is the usual time-independent test statistic. Thus, in the fine-binned limit, the sensitivity to a time-dependent signal is significantly enhanced if systematic uncertainties dominate over statistical ones.  The time-dependence of the signal allows the background rate to be measured 
in situ, in the low signal rate bins, thus removing the associated uncertainty. In fact, in the systematics-dominated regime, Eq.~\eqref{eq:finebinning-main} reduces to~\cite{SuppMat}
\be
\chi^2_{\rm fb} =  \frac{({\cal B}_{\rm sig} f_{\rm sig} N_{\rm tot})^2}{2 \,n_{\rm bin} \,\sigma^2_{\rm stat}},
\ee
which shows complete independence from systematic uncertainties. Thus, with an appropriate analysis, it is possible to measure a time-dependent
signal even in the presence of large systematic uncertainties, as long as all the systematics are time-independent.

The expected sensitivity generally depends on the choice of time-bin width $\Delta t$ (equivalently, the number of bins $n_{\rm bin}$), as shown in Fig.~\ref{fig:upper_limit_vs_delta_t}. The time-independent limit corresponds to $n_{\rm bin}=1$ (horizontal dashed line in Fig.~\ref{fig:upper_limit_vs_delta_t}). Sensitivity gradually decreases as $\Delta t$ increases: small $\Delta t$ corresponds to a regime fully resolving oscillations, whereas large $\Delta t$ approaches a less sensitive, time-integrated analysis dominated by systematics.  

{The $\chi^2$ approach requires at least 
$\sim10$ events per time bin which places a lower bound on the bin size $\Delta t$.  To resolve shorter period oscillations, but above the experimental time resolution, we utilize the unbinned approach of the Rayleigh periodogram \cite{Losada:2023zap}.  Further details on the procedure can be found in the Supplemental Material~\cite{SuppMat}.  In Fig.~\ref{fig:test}, we summarize the expected sensitivity to ${\cal B}(\ell_i\to \ell_j\phi')$ as a function of $m_\phi$ for representative numerical examples detailed below.}

\medskip
\noindent{\bf Numerical examples.} 
As representative examples we consider a possible search for time-dependent $\mu\to e\phi'$ decays at Mu3e, and for $\tau\to \ell\phi'$ decays at Belle II and at FCC-ee. More details regarding the chosen numerical values for each example can be found in the Supplemental Material~\cite{SuppMat}.

For Mu3e, we assume $T = 300$ days of continuous data collection, yielding $N_{\rm tot}\sim 3 \times 10^{15}$ muon decays. The endpoint region for $m_\phi \simeq 0$ includes $N_{\rm bg}\sim 10^{13}$ background events from $\mu\to e \nu\bar \nu$, corresponding to a background fraction $f_{\rm bg}\sim 3 \times 10^{-3}$~\cite{Perrevoort:2018okj, Perrevoort:2023qhn}.
 For these large data samples, the $\mu\to e\phi'$ search will be dominated by systematics, which we conservatively assume are set by theoretical uncertainties, implying a 90\% CL upper bound of ${\cal B}(\mu\to e\phi')< 6 \times 10^{-7}$~(Fig.~12 of Ref.~\cite{Banerjee:2022nbr}), and thus a relative systematic uncertainty of $\alpha\sim 10^{-4}$. 
Given the dataset above, the time-dependent search for $\mu\to e\phi'$ can result in a sensitivity to the $\mu\to e\phi'$ branching ratio well below the systematic uncertainty for $m_\phi\lesssim 10^{-10}\,$eV (solid purple line in Fig.~\ref{fig:test} left).

Estimating the Belle-II sensitivity to time-dependent $\tau\to \ell\phi'$ decays is complicated by the fact that the state-of-the-art time-integrated search for these decays relies on the combined fit to the shape of $\tau \to \ell\nu\bar\nu$ background  and the $\tau\to \ell\phi'\,$  signal in the pseudo-rest-frame of the tau~\cite{Belle-II:2022heu}. Since, in this frame, the spectrum of lepton energies in  SM $\tau \to \ell\nu\overline{\nu}$ decays is similar to the spectrum from $\tau \to \ell\phi'$, we approximate the sensitivity using a simplified counting experiment. 
For the existing Belle-II analysis \cite{Belle-II:2022heu}, using data corresponding to an integrated luminosity of 62.8 fb$^{-1}$, we consider only the $\tau\to \ell \nu\overline{\nu}$ events with lepton energies in the range that contains $90\%$ of the signal leptons from $\tau \to \ell \phi'$ decays. Taking experimental efficiencies into account gives $f^{e[\mu]}_{\rm sig} \sim 1.8[2.4]\times 10^{-2}$ 
and $f^{e[\mu]}_{\rm bg} \sim 1.1[1.4]\times 10^{-2}$.
Using this, the background process branching $\mathcal{B}^{e[\mu]}_{\rm bg} \sim 0.178[0.174]$, the 90\% CL upper limits set by the analysis $B(\tau \to e[\mu] \phi') < 7.6[4.7] \times 10^{-4}$~\cite{Belle-II:2022heu}, and assuming that the search is systematics limited,  gives the relative systematic uncertainty of $\alpha_{e[\mu]} \sim 2.7[1.9]\times 10^{-2}$.
For simplicity, we also assume that the $N_{\rm tot}\sim 10^8$ taus were collected uniformly within $T=300$~days. Using this relatively small data-sample, the time-dependent search resolves oscillations only for $\phi$ masses below $m_\phi \lesssim 10^{-17}$\,eV (blue solid line in Fig.~\ref{fig:test} right).
With the full 50 ab$^{-1}$ integrated luminosity, Belle-II will have $N_{\rm tot} \sim 10^{11}$, which should significantly improve the reach in $f^2/C_{\tau \ell}$ and sensitivity to larger $\phi$ masses. Assuming continuous data collection over $T = 3\times10^{3}$ days with unchanged $f_{\rm bg}$, $f_{\rm sig}$, and $\alpha$, yields the blue dashed-dotted line in Fig.~\ref{fig:test} (right).

Finally, we also consider $\tau\to \ell\phi'$ searches at FCC-ee. While running in Tera-Z mode the proposed FCC-ee experiment will produce 
approximately $N_{\rm tot}=3.4\times 10^{11}$ taus over $T=740$ days of running \cite{Bernardi:2022hny}. Using a relative systematic uncertainty of $\alpha \sim 2.6 \times 10^{-4}$, $f_{\rm bg} \sim 8.1 \times 10^{-2}$, and $f_{\rm sig} \sim 0.12$ we obtain the projected sensitivities shown as the red solid line in Fig.~\ref{fig:test} (right), resolving oscillations up to $m_\phi \lesssim 10^{-13}$ eV.

\medskip

\noindent{\bf Astrophysical and laboratory
constraints.}
Light particles with CFLV interactions can modify energy transport in supernovae and conflict with the observed properties of SN1987A in the Large Magellanic Cloud (LMC) ~\cite{Calibbi:2020jvd}.  Previous bounds on such particles assumed environment-independent couplings, but our interaction from Eq.~\eqref{eq:Lint:main2} depends on $\phi_c$ and thus requires knowledge of the DM density in the LMC. At the SN1987A location ($\sim 1$ kpc from the LMC center \cite{Panagia:1991}), the DM density is estimated to be  \cite{Buckley:2015doa}
 \be
 \rho^{\rm LMC}_\phi\in[0.4,8]\,\text{GeV}/\text{cm}^{3}~,
\ee
and in Fig.~\ref{fig:test} (left) we use the lower end of this range to place conservative limits on the quantity $f/\sqrt{C_{\mu e}} $, following the same procedure as outlined in Ref.~\cite{Calibbi:2016hwq}.

It is instructive to compare CLFV constraints to those on flavor-diagonal couplings. 
As diagonal scalar couplings vanish in our model at leading order, see Eq.~\eqref{eq:Lint:main2}, we consider constraints on an ultralight pseudoscalar DM and define ${C_{\mu e}}|_{\text{Mu3e}}$ as the maximum Mu3e sensitivity in Fig.~\ref{fig:test}, reflecting a time-binned analysis with fine binning. 
The strongest constraints on the diagonal electron couplings are from Red Giant cooling~\cite{Capozzi:2020cbu} 
\be
\label{eq:Cee}
   C_{ee}\left(\frac{\text{GeV}}{f}\right)^2 \lesssim 2
   \times 10^{2} \left(\frac{m_\phi}{\text{eV}} \right),
\ee
while for the diagonal muon coupling, the strongest constraints come from SN1987A cooling, where ~\cite{Bollig:2020xdr, Caputo:2021rux}
\be
C_{\mu\mu}\left(\frac{\text{GeV}}{f}\right)^2 \lesssim 8\times 10^{4}  \left(\frac{m_\phi}{\text{eV}} \right),
\ee
which yields the relations
\be \label{eq:diagonalP}
C_{ee}\lesssim 30\, C_{\mu e}|_\text{Mu3e}~,~C_{\mu\mu}\lesssim  10^4
\, C_{\mu e}|_\text{Mu3e}.
\ee
Without fine-tuned cancellations, rotating into the charged-lepton mass basis generally produces
\be 
 C_{\mu e}\lesssim \sqrt{C_{ee}C_{\mu\mu}} \lesssim 5\times 10^2\, C_{\mu e}|_\text{Mu3e}
\ee
where we have combined Eqs.~\eqref{eq:Cee}$-$\eqref{eq:diagonalP} in the last expression. This indicates that Mu3e searches can probe new and \textit{natural} DM parameter space without special hierarchies between diagonal and off-diagonal couplings.
Given the much weaker bounds on $C_{\tau\tau}$, the Belle-II searches we describe above also do not face any such fine-tuning considerations, especially for $\tau \to\mu \phi'$ decays.

\medskip

\noindent{\bf Conclusions.}
In this {\it Letter} we demonstrated that FCNC decays may offer a new window into ultralight DM candidates.
Crucially, the presence of FCNC couplings to charged SM leptons can induce time modulation with DM mass frequency in $\tau\to \ell \phi'$ and  $\mu\to e\phi'$ decays, where $\phi'$ is a light dark sector scalar. Since observing time dependence in these decays would be smoking-gun evidence of DM, we find that rare lepton decay experiments can also serve as potential dark matter direct detection experiments without any instrumental modifications. Furthermore, such time-oscillating signals can improve the experimental reach compared to time-independent searches, if the latter are dominated by systematic uncertainties, see Fig.~\ref{fig:test}. 

While our analysis focused on flavor violation in charged lepton decays, the same approach directly extends both to other observables in the lepton sector (such as $\mu \to e$ conversion in the field of a nucleus), as well as to flavor violation in the quark sector. We leave a more detailed analysis of these phenomena for future work.

\vspace{-0.1in}
\begin{acknowledgments}
\vspace{-0.1in}
\noindent 
We are indebted to Joachim Kopp for physics discussions during the early stages of this work, and to Gilad Perez for enlightening discussions about naturalness in set-ups similar to ours.  This manuscript has been authored in part by  Fermi Forward Discovery Group, LLC under Contract No.~89243024CSC000002 with the U.S.~Department of Energy, Office of Science, Office of High Energy Physics. The work of IB, PF and RH was performed in part at the Aspen Center for Physics, supported by a grant from the Alfred P. Sloan Foundation (G-2024-22395). IB, PF, RH and JZ  were supported in part by the Munich Institute for Astro-, Particle and BioPhysics (MIAPbP) which is funded by the Deutsche Forschungsgemeinschaft (DFG, German Research Foundation) under Germany´s Excellence Strategy – EXC-2094 – 390783311. YG acknowledges support by the Cluster of Excellence ``PRISMA+'' funded by the German Research Foundation (DFG) within the German Excellence Strategy (Project No. 390831469), and by a fellowship awarded by the Azrieli Foundation.  JZ and TM acknowledge support in part by the DOE grant de-sc0011784, and NSF grants OAC-2103889, OAC-2411215, and OAC-2417682. JZ and TM also acknowledge support in part from the Visiting Scholars Award Program of the Universities Research Association.
\end{acknowledgments}

\onecolumngrid
\bibliography{letter_main}
\newpage 

\appendix
\section*{Supplemental Material}
\renewcommand{\theequation}{S\arabic{equation}}
\setcounter{equation}{0}

\subsection{Non-Abelian pseudo-NGBs} 
\label{sec:npNGBs}

A spontaneously broken global symmetry gives rise to massless Nambu-Goldstone Bosons (NGBs). The NGBs acquire a nonzero mass, if there is a small explicit breaking of the global symmetry. Celebrated examples of NGBs in particle physics are the QCD axion, and the light mesons $\pi^\pm$ and $\pi^0$. The QCD axion is a pseudo-NGB (pNGB) of a $U(1)_{\rm PQ}$ Peccei-Quinn global symmetry, spontaneously broken at high scale $f_a$ and then also explicitly 
broken by the QCD anomaly \cite{Peccei:1977hh,Peccei:1977ur,Wilczek:1977pj,Weinberg:1977ma}. The pions are the pNGBs of spontaneous chiral symmetry breaking in QCD, $SU(2)_L\times SU(2)_R\to SU(2)_V$, where the initial $SU(2)_L\times SU(2)_R$ global group is also explicitly broken by the $m_{u,d}$ quark masses (see, e.g., \cite{Donoghue:1992dd}).
The QCD axion is an example of an Abelian pNGB, while the pions are examples of non-Abelian pNGBs. 

More generally, the non-Abelian pseudo-Goldstone Bosons (npNGBs) arise when the theory is invariant
under an approximate non-Abelian global  group $G$, that is  spontaneously broken to its subgroup $H$ ($G\to H$), where the $G/H$ coset  consists of several pNGBs that have non-linear interactions with each other.  Let us denote the npNGB parametrization of the  $G/H$ coset as $U(\phi)$ where, under $G$ transformations, $U(\phi)\to V_L^\dagger U(\phi) V_R$. In general, the left and right transformations differ so that $V_L\ne V_R$; this is the case we are interested in.

We assume that the leading interactions between the npNGB dark sector and the SM has the general form 
\beq
\label{eq:UdU:current}
{\cal L}_{\rm int}\supset \Tr\big(Q U^\dagger i\partial_\mu U\big) \bar \ell_i \gamma^\mu (\tilde C^V_{\ell_i \ell_j} + \tilde C^A_{\ell_i \ell_j} \gamma_5 ) \ell_j +{\rm h.c.}, 
\eeq
where $\ell_i$ are the SM charged leptons, and $Q$ is the symmetry breaking spurion  that transforms as $Q\to V_R^\dagger Q V_R$ under $G$. Other transformation properties for $U$ and $Q$, and thus different structures for the interaction term, Eq. \eqref{eq:UdU:current}, can also be considered. The interaction in Eq.~\eqref{eq:UdU:current} can be generated via heavy mediators, that are then integrated out at low energies, but 
variants of this model with light mediators can also be considered (for the phenomenology we are most interested in, though, they should still be heavier than a muon).  Note that the interaction term Eq.~\eqref{eq:UdU:current} remains invariant under shift symmetry, $U\to e^{-i \alpha} U $, where $\alpha=\alpha^a T^a$ is an arbitrary constant matrix in the tangent space of the $G/H$ coset. Due to this remaining shift symmetry the interaction in Eq.~\eqref{eq:UdU:current} does not generate contributions to npNGB masses.

Expanding the $U(\phi)$ exponentiation in terms of the $\phi_a$ fields, $U=\exp(i \phi^a T^a)$, gives 
dimension-6 interaction operators with a typical form 
\beq
\label{eq:Lint:nonabel}
{\cal L}_{\rm int}\supset  \Big(\frac{\phi_1}{f} \frac{i\partial_\mu \phi_2}{2 f} - \frac{\phi_2}{f} \frac{i\partial_\mu \phi_1}{2 f}\Big)\bar \ell_i \gamma^\mu (C^{V}_{\ell_i \ell_j} + C^{A}_{\ell_i \ell_j} \gamma_5 ) \ell_j +{\rm h.c.}, 
\eeq
where $f$ is the scale of spontaneous symmetry breaking, while $\phi_1$ and $\phi_2$ are two different npNGBs, possibly related through complex conjugation so that $\phi_1=\phi_2^\dagger$, depending on the detailed structure of the $G/H$ coset and on the form of spurion $Q$. Note that the above interaction term is invariant under the nonlinearly realized shift symmetry $\phi^a T^a \to \phi^a T^a-\alpha^a T^a -\frac{1}{2}[\alpha^a T^a, i\phi^a T^a]+\cdots$, with $\alpha^a$ constants (this is most easily seen in the exponentiated form, keeping all terms in $\phi^a$).

It is illuminating to compare the phenomenology of the FCNC $\ell_i\to \ell_j \phi$ decays for the two cases, when $\phi$ is a non-Abelian pNGB with interaction of the form \cref{eq:Lint:nonabel}, and when $\phi$ is an Abelian pNGB with an interaction that involves only a single $\phi$ field, 
\beq
\label{eq:Lint:pNGB}
{\cal L}\supset  C_{\ell_i \ell_j} \brac{\partial_\mu \phi}{f}   \overline \ell_i \gamma^\mu  \gamma_5  \ell_j.
\eeq
There are several important differences between the two cases. First of all, since the interaction in Eq.~\eqref{eq:Lint:pNGB} is linear in $\phi$, the $\ell_i\to \ell_j \phi$ decay rates for Abelian pNGB case are time independent, and also do not depend on the local DM number density. In contrast, 
for npNGBs in \cref{eq:Lint:nonabel} the two-body $\ell_i\to \ell_j \phi$ decays arise only, if there is a background density of light $\phi$ particles. This background density then leads to time-dependent $\ell_i\to \ell_j \phi$ rates. The observation of such time-dependent $\ell_i\to \ell_j \phi$ decays would be a smoking-gun signal of light npNGB dark matter. For Abelian pNGB the observation of $\ell_i\to \ell_j \phi$ signal instead does not immediately imply that $\phi$ is DM, and one would need to confirm that $\phi$ is indeed the DM using other observations (for instance by searching for time-dependent interactions via its couplings to electrons).

 The npNGBs have already been discussed in the literature as the possible dark matter candidates, though, in a very different mass regime that we are interested in. An example is the strongly interacting massive particle (SIMP) dark matter candidate \cite{Hochberg:2014kqa,Hochberg:2015vrg,Choi:2017zww,Hochberg:2018rjs,Hochberg:2018vdo}. While in the SIMP case the interesting mass regime is around GeV, we are interested in much lighter npNGB dark matter candidates, with masses well below eV.  For other examples of heavier npNGB dark matter, see, e.g., Refs. \cite{Krnjaic:2014xza,Balkin:2018tma,Balkin:2017yns,Marzocca:2014msa,Frigerio:2012uc}, and for strongly-interacting dark sectors with dark pions as npNGBs, see, e.g., Refs. \cite{Berlin:2018tvf,Bernreuther:2019pfb,Kribs:2018ilo,Renner:2018fhh}. The dark sector currents of the form $\phi\partial_\mu \phi$ coupling to SM fermion currents through light dark vectors were also considered in \cite{Eguren:2024oov}.

\subsubsection{An example realization: $SU(3)_L\times SU(3)_R\to SU(3)_V$} 

The form of the interactions in Eq.~\eqref{eq:UdU:current} may seem exotic. However, as noted above, there is a familiar example in the SM: the pion sector of QCD interacting through QED with leptons.  In the limit of vanishing up and down quark masses the QCD Lagrangian is invariant under a chiral flavor symmetry, $SU(2)_L\otimes SU(2)_R$. This is spontaneously broken to its diagonal subgroup, $SU(2)_L\otimes SU(2)_R\to SU(2)_V$, resulting in three light npNGBs, $\pi^\pm$ and $\pi^0$. In the QCD Lagrangian the  $SU(2)_L\otimes SU(2)_R$ global symmetry is explicitly broken by the $m_{u,d}$ quark masses, giving rise to nonzero pion masses. The $\pi^+ i \partial_\mu \pi^-$ current  couples to the electromagnetic current of the SM leptons through a tree level photon exchange, giving rise to a non-local interaction of the form $\big(\pi^+ i \partial_\mu \pi^-\big)\partial^{-2}\big(\bar \ell \gamma^\mu \ell\big)$, where the nonlocal structure $\partial^{-2}$ is due to the photon propagator, $-i g_{\mu\nu}/q^2$, but in the position-space representation. 

If all three light SM quark masses, $m_u,m_d$ and $m_s$ can be neglected, then the QCD Lagrangian has a larger global symmetry $SU(3)_L\otimes SU(3)_R$. This can still be viewed as being spontaneously broken to its diagonal subgroup $SU(3)_V$, though, in this case the explicit breaking due to strange quark mass $m_s$ is much larger. In this case there are eight npNGBs: the pions, kaons and eta. 

Let us now assume that a similar $SU(3)_L^\text{hid}\otimes SU(3)_R^\text{hid}\to SU(3)_V^\text{hid}$ symmetry breaking pattern also occurs in the hidden sector. 

Let us also charge the dark sector under a gauged $U(1)'$. As in the introductory discussion surrounding \cref{eq:UdU:current}, let us assume that $U(1)'$ is a subgroup of $SU(3)_R^\text{hid}$. 

The only explicit breaking of $SU(3)_L^\text{hid}$ are thus the masses of hidden sector quarks, which guarantees that the $\pi_D^a$ can be very light if the hidden quark masses are small. 
-- that may, but need not, be a subgroup of $SU(3)_V^{\rm hid}$, and assume that $\pi_D^\pm$ are the dark matter. That is, the low energy chiral Lagrangian for the hidden sector is given by
\beq
\label{eq:L:dark:chpt}
{\cal L}=\frac{f_{\rm UV}^2}{8}\Tr\big(D_\mu U D^\mu U^\dagger\big)+\frac{f_{\rm UV}^2}{8}\Tr\big(\chi^\dagger U +  U^\dagger \chi \big)+\cdots, 
\eeq
where $\chi$ is an explicit $SU(3)_L^\text{hid}\otimes SU(3)_R^\text{hid}$ global symmetry breaking spurion due to hidden quark masses $m_q^\text{hid}$ (we use the notation similar to QCD, thus $\chi\propto (m_u^\text{hid},m_d^\text{hid},m_s^\text{hid})$). The ellipses in \cref{eq:L:dark:chpt} denote  terms of higher order in chiral expansion. Working in the basis where $\chi$ is diagonal, 

the exponentiated npNGB matrix  $U=\exp\big(i \sqrt{2} \Pi_D/f_{\rm UV}\big)$ is given by
\beq
\Pi_d=
\begin{pmatrix}
\frac{\pi_D^0}{\sqrt{2}} + \frac{\eta_{D8}}{\sqrt{6}} & \pi_D^+ & K_D^+ 
\\
 \pi_D^- & -\frac{\pi_D^0}{\sqrt{2}} + \frac{\eta_{D8}}{\sqrt{6}}  & K_D^0 
\\
K_D^-  & \bar{K}_D^0 & -\sqrt{\frac{2}{3}}\eta_{D8}
\end{pmatrix},
\eeq
where again we follow the SM notation for the npNGB mass eigenstates. (Note that the charges in this notation need not correspond to $A'$ charges.) 
The covariant derivative is given by
\beq
D_\mu U=\partial_\mu U+ig' A_\mu' Q U,
\eeq
where $g'$ is the $U(1)'$ gauge coupling, and $Q$ is a $3\times 3$ spurion matrix. For instance, taking $Q=q_U \lambda_3$, where $\lambda_3$ is the third Gell-Mann matrix, and $q_U$ the appropriately normalized $U(1)'$ charge, the kinetic term contains
\beq
\label{eq:lambda6}
\begin{split}
\frac{f_{\rm UV}^2}{8}\Tr\big(D_\mu U D^\mu U^\dagger\big) & \supset\frac{f_{\rm UV}^2}{8}g' A_\mu' \Tr\Big(QUi\partial_\mu U^\dagger -i\partial_\mu U U^\dagger Q\Big)+\cdots
\\
&= \frac{f_\text{UV} g' q_u}{2\sqrt{2}} A_\mu' \partial_\mu \pi_D^0   +\frac{g' q_U}{4} A_\mu' \big(\bar K_D^0 i\partial^\mu K_D^0- K_D^0i\partial^\mu \bar K_D^0\big)+\cdots,
\end{split}
\eeq
where ellipses denote terms with no gauge fields or with two gauge fields, and the terms where $A'$ couples to currents with three or more mesons, or to two mesons, but not involving $K_D^0, \bar K_D^0$.

Dark meson masses are proportional to dark quark masses, $m_{\pi_D^0}^2\propto (m_u+m_d)$, $m_{K_D^0}^2=m_{\bar K_D^0}^2\propto (m_s+m_d)$, and so on, completely analogous to QCD. For $m_u\gg m_{d,s}$  the $K_D^0$ and $\bar K_D^0$ are much lighter than all the npNGBs, and can be the DM candidates. We are interested in the regime, where   $K_D^0$ and $\bar K_D^0$ are extremely light and are the DM, with the relic abundance set by the misalignment mechanism, while all the other npNGBs are heavier than the muon and thus irrelevant for muon decay phenomenology. 

We can also charge the SM fermions under this dark $U(1)'$, in principle with flavor violating couplings, see, e.g., \cite{Smolkovic:2019jow}. In this example we do not aim to explain the structure of SM Yukawas, and so interactions will have a general flavor structure
\beq
{\cal L}_{\rm int}=i g' c_{ij} \bar \psi_i \slashed A' \psi_j +{\rm h.c.}.
\eeq
We set  $Q=q_U \lambda_3$ as in \cref{eq:lambda6}, and take the mass of the dark photon to be generated via a dark-Higgs mechanism, with vacuum expectation value $v'$, such that $m_{A'}=g'v'$.
The tree-level exchange of $A'$ between dark sector and SM fermions generates an interaction of the form

\beq
c_{ij} \frac{ g'{}^2 f_{\rm UV}^2}{8m_{A'}^2} \big(\bar \psi_i \gamma^\mu \psi_j) \Tr\Big(QU i \partial_\mu U^\dagger -i\partial_\mu U U^\dagger Q\Big)= \frac{c_{ij} q_U g'^2 }{4 m_{A'}^2}  \big(\bar \psi_i \gamma^\mu \psi_j\big)  \big(\bar K_D^0 i\partial_\mu K_D^0 +{\rm h.c.}\big) +\cdots.
\eeq
Note that the $f_{\rm UV}$ scale disappears from the effective interaction. The strength of the interaction with the SM is instead controlled by $m_{A'}/g'=v'$, i.e., by the $U(1)'$  spontaneous symmetry breaking scale.

 Matching onto the notation in Eq.~\eqref{eq:Lint:nonabel} gives\footnote{The correspondence with the example in the main text is not exact, since we have DM that is composed out of two states, $K_D^0$ and $\bar K_D^0$.}
\beq\label{eq:appfrelation}
\frac{f}{\sqrt{C_{ij}}} \sim \frac{2 m_{A'}}{g'\sqrt{c_{ij}q_U}} = \frac{2\cdot 10^9 \,\text{GeV}}{\sqrt{c_{ij}q_U}} \biggr(\frac{m_{A'}}{10^6\text{\,GeV}}\biggr)\biggr(\frac{10^{-3}}{g'}\biggr).
\eeq

\subsubsection{Other examples} 
Beyond npNGBs there are other possibilities for quadratic interactions between dark sector and the SM. We are interested in the case where there are two different species of dark sector particles involved in the couplings to the SM currents (for the case where a single light DM field couples quadratically to the SM with flavor diagonal interactions,  see, e.g., \cite{Banerjee:2022sqg,Gan:2025nlu,Grossman:2025cov,VanTilburg:2024xib,Beadle:2023flm,Hees:2018fpg}). We are also specifically interested in the CLFV SM currents. 
The simplest such interaction is, for instance, 
\beq
\label{eq:scalar:int}
  \left(\frac{\phi\, \phi^\prime}{\Lambda_{ij}^2}\right) H \overline{L}_i \ell _j.
\eeq
Linear interaction of $\phi$ and $\phi'$ may be forbidden, if these are odd under a $Z_2$ symmetry. 
The main difference with respect with the npNGBs e is that the masses of $\phi$ and $\phi's$ are no longer protected using just the symmetries still present in the IR. Such interactions may  thus face a large hierarchy problem, though, this may not be a problem in concrete UV models, e.g,.~if the dark sector possesses a softly broken supersymmetry.

Another example where multiple dark sector particles may interact with the SM current is if the dark matter is composed of light non-Abelian gauge bosons, $G_{D\mu}^a$, where the gauge symmetry is spontaneously completely broken.  These dark gauge bosons can interact with the SM through a higher dimension operator such as 
\beq
\label{eq:non-Abel}
  \frac{1}{\Lambda_{ij}^4}\big(G_{D\mu\nu}^a G_{D}^{a\mu\nu}\big) \big( H \overline{L}_i \ell _j\big)
  ~,
\eeq
with $G^a_{D\mu\nu}$ the field strength associated with the vector field $G_{D\mu}^a$.
These interactions contain terms that are of the schematic form $\partial G_D \partial G_D \bar \mu e$, $\partial G_D G_D^2  \bar \mu e$, and $G^4 \bar \mu e$. The $\partial G_D \partial G_D \bar \mu e$ operator is not interesting for our purposes since it leads to effects that are suppressed by the small $G_D$ mass. The other two operators, on the other hand, can lead to $\mu \to e G_D$ decays with either two or three $G_D$ fields set to their classical background values. The $\mu \to e G_D$ decay rates will thus oscillate as a combination of higher harmonics $\Gamma(\mu \to e G_D)\propto\big(\cos^2 m_{G_D}t +\ldots \cos^3 m_{G_D}t\big)^2$. The operator in \cref{eq:non-Abel} can also induce time dependent $\mu\to e G_D G_D$ decays, which can also be searched for.

\subsection{\texorpdfstring{Time-dependent $\chi^2$ analysis}{Time-dependent analysis techniques} }

Here we present the details necessary to estimate the time-dependent sensitivity  projection for the sinusoidal $\ell_i \to \ell_j \phi'$ signal from Eq.~\eqref{eq:li:to:lj:phi}.  To simplify the analysis we consider an idealized case: an experiment that can collect data continuously for a total observation time $T > \tau_\phi$, with a constant time resolution $\Delta t < \tau_\phi$, and with a constant efficiency close to unity. In a realistic experimental analysis these simplifying assumptions may not be realized, implications of which we will discuss in Sec.~\ref{sec:result}. 

Since galactic DM is non-relativistic with local velocity ${\mathcal O}(v)\sim 10^{-3}$, its energy spread is ${\mathcal O}(v^2)\sim 10^{-6}$, so the 
the signal in Eq.~\eqref{eq:li:to:lj:phi} is coherent over ${\mathcal O}(10^6)$ oscillations. Therefore, in our analysis we separately consider two cases: 
\begin{enumerate}
    \item {\bf Coherent (``slow'') oscillations:} for sufficiently slow oscillations, we can assume that the signal is coherent over the complete data collection time, which we discuss in Sec.~\ref{sec:single:coherent},
    \item {\bf Incoherent (``fast'') oscillations:}  for faster oscillations we need to take into account the random phase variations using $\delta$ in Eq.~\eqref{eq:li:to:lj:phi}, as discussed in Sec.~\ref{sec:multiple:patches}.
\end{enumerate}
{The time-integrated analysis is given in Sec.~\ref{sec:time-indep:signal} and an unbinned analysis, using the Rayleigh periodogram, in Sec.~\ref{sec:Rayleigh}.
}

\subsubsection{Coherent (``slow'') oscillations}
\label{sec:single:coherent}

{
We first consider time-oscillating $\ell_i \to \ell_j \phi'$ decays with a fixed phase and a period that is long enough it can be resolved with a binned analysis.  We later consider the effects of phases varying across multiple coherence patches (Sec.~\ref{sec:multiple:patches}).}
To estimate the expected upper bound on the branching ratio of the oscillating signal, we construct a $\chi^2$ test statistic, and work with the ``Asimov" data set, which assumes that each of the $n_{\rm bin}=T/\Delta t$ bins will contain the ``predicted'' event counts ($N_\text{pred}$)
under a given hypothesis. Since background processes are assumed not to oscillate, the expected number of background events is taken to be constant in each bin. Thus, in the presence of the oscillatory signal from Eq.~\eqref{eq:Ndot} with fixed phase $\delta$, the total number of predicted number of events in $k^\text{th}$ time bin, with $t\in[t_{k-1}, t_{k}]$, is given by  
\beq
\begin{split}\label{eq:oschypo}
  N_{\text{pred},k} =\int_{t_{k-1}}^{t_{k}} dt \dot{N}_{\text{pred}} =\frac{N_{\rm tot}}{T}  \int_{t_{k-1}}^{t_{k}} dt \left[\mathcal{B}_{\rm bg} f_{\rm bg} +  2\mathcal{B}_{\rm sig} f_{\rm sig} \cos^2 (m_\phi t + \delta) \right]  , 
\end{split}
\eeq
where $N_{\rm tot}$ is the total number of observed $\ell_i$ decays, 
$\mathcal{B}_{\rm sig}$  ($\mathcal{B}_{\rm bg}$) is the $\ell_i \to \ell_j \phi'$ ($\ell_i \to \ell_j\nu\bar \nu)$ \textit{time-averaged} branching ratio, while $f_{\rm{sig}(\rm{bg})}$ is the experimental efficiency -- which includes both detection efficiencies and experimental cuts. 
As anticipated, under the background-only hypothesis, the expected number of observed events is constant across all bins, so 
\beq
N_{{\rm bg},1} \equiv N_{\text{pred},k} \bigr|_{{\cal B}_{\rm sig} \to 0} =   N_{\rm tot} \mathcal{B}_{\rm bg} f_{\rm bg}  \left( \frac{ \Delta t}{T} \right),
\eeq
where we have defined $N_{{\rm bg},1}$ to be the constant background count in each bin.
In the limit of large backgrounds, we take the statistical uncertainty in each bin to satisfy $\sigma_{\rm stat}^2 = N_{{\rm bg},1}$, consistent with the use of the Asimov set. We also take into account systematic uncertainties, modeled by $\sigma_{\rm sys}^2 = \alpha^2 N_{{\rm bg},1}^2$ with $\alpha$ common to all bins -- i.e., we assume that the systematic error is a fixed, time-independent, relative error. Therefore, systematics are fully correlated across all time bins, and the covariance matrix and its inverse are given by
\beq
  C = N_{{\rm bg},1} \mathbf{1} + \alpha^2 N_{{\rm bg},1}^2 \mathbb{I}, \quad \textrm{and}\quad  C^{-1} = \frac{1}{N_{{\rm bg},1}} \mathbf{1} - \frac{\alpha^2}{1 + \alpha^2 N_{{\rm bg},1} n_{\rm bin}} \mathbb{I},
\eeq
where $\mathbf{1}$ is the $n_{\rm bin} \times n_{\rm bin}$ identity matrix and $\mathbb{I}$ is the $n_{\rm bin} \times n_{\rm bin}$ constant unity matrix with $1$ in every entry. The corresponding $\chi^2$ statistic is therefore given by 
\beq \label{eq:chi2}
\begin{split}
\chi^2 &= \sum_{k,p = 1}^{n_{\rm bin}} S_k C^{-1}_{kp} S_p= \frac{1}{N_{{\rm bg},1}}\sum_{k = 1}^{n_{\rm bin}} S_k^2 - \frac{\alpha^2}{1 + \alpha^2 n_{\rm bin} N_{{\rm bg},1}}\left( \sum_{k = 1}^{n_{\rm bin}} S_k\right)^2~,
\end{split}
\eeq
where $S_k \equiv N^k_{\rm obs} - N_{\rm pred}^k$ is the number of signal events in bin $k$. For the signal in Eq.~\eqref{eq:oschypo} we then have
\beq \label{eq:int_signal}
\begin{split}
  S_k &= 2 \mathcal{B}_{\rm sig} f_{\rm sig} \frac{N_{\rm tot}}{T} \int_{(k-1)\Delta t}^{k\Delta t} dt \cos^2(m_\phi t + \delta) \\
  &=  \mathcal{B}_{\rm sig} f_{\rm sig} \frac{N_{\rm tot}}{T} \left[ \Delta t + \frac{\sin\left(2k m_\phi \Delta t + 2 \delta\right) - \sin\left(2(k-1)m_\phi \Delta t +2 \delta\right)}{2 m_\phi} \right],
\end{split}
\eeq
\begin{figure*}[t!]
\includegraphics[width=0.49\textwidth]{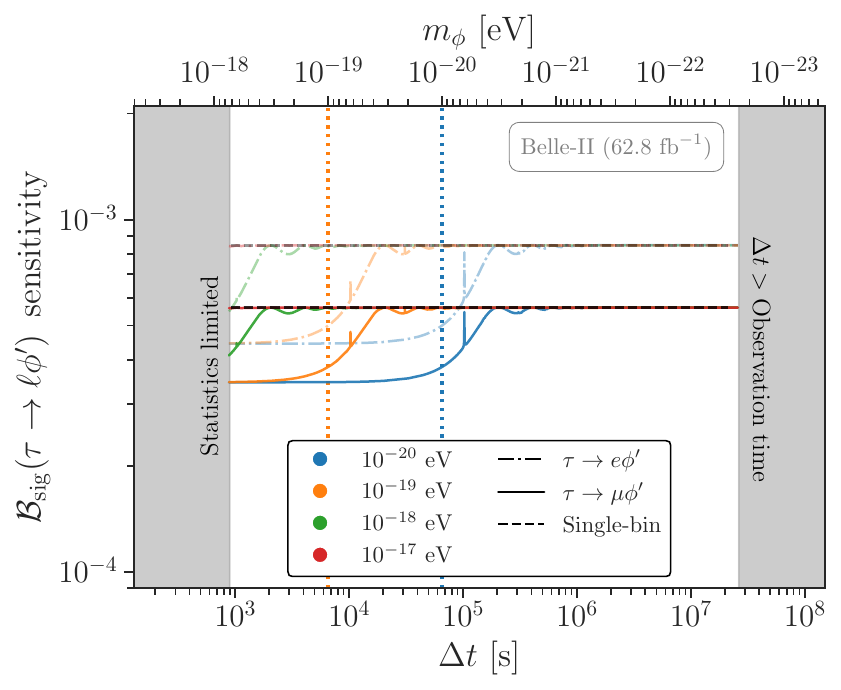}
\includegraphics[width=0.49\textwidth]{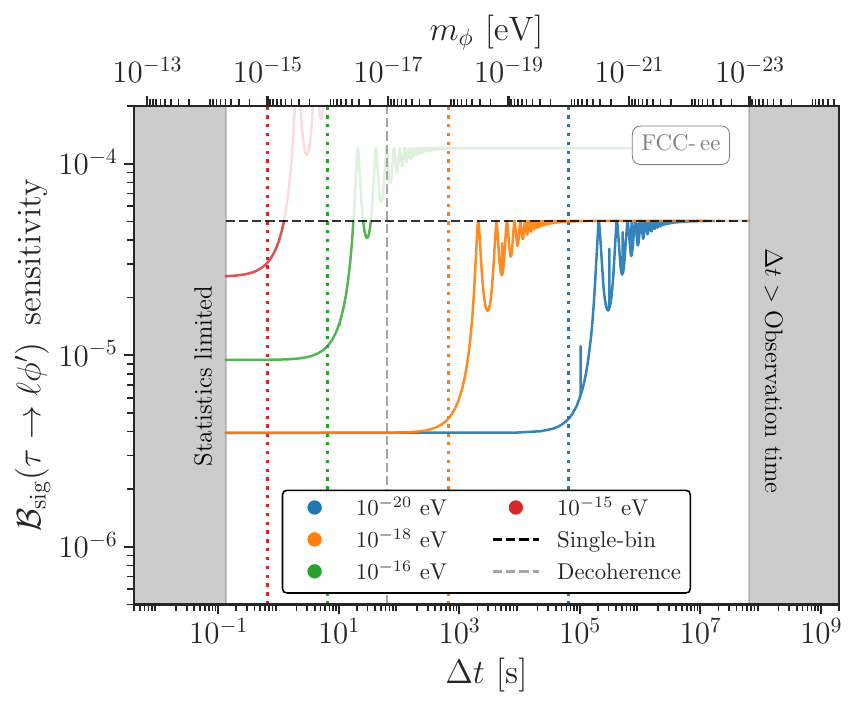}
  \caption{ Same as Fig.~\ref{fig:upper_limit_vs_delta_t} for the current integrated luminosity at Belle-II (left) as well the projection for FCC-ee (right).
  }
  \label{fig:upper_limit_vs_delta_t_BelleII}
\end{figure*}
so using this expression, the two terms in the $\chi^2$ in Eq.~\eqref{eq:chi2} can be computed explicitly,  and satisfy
\be
\label{eq:sum_S_i}
\sum_{k = 1}^{n_{\rm bin}} S_k &=& \mathcal{B}_{\rm sig} f_{\rm sig} N_{\rm tot} \left[ 1 + \frac{\sin(2 m_\phi T + 2 \delta) - \sin(2 \delta)}{2m_\phi T} \right],
\\
 \label{eq:sum_S^2_i}
\sum_{k = 1}^{n_{\rm bin}} S_k^2 &=& \mathcal{B}_{\rm sig}^2 f_{\rm sig}^2 \left(\frac{N_{\rm tot}}{2 m_\phi T}\right)^2 \biggl\{ n_{\rm bin} \big[1 + 4 \Delta t^2 m_{\phi}^2 - \cos(2 m_\phi \Delta t)\big] 
\nonumber\\
&& \hspace{1cm}
+ 8m_{\phi} \Delta t \sin(m_\phi T)  \cos( m_\phi T + 2\delta)
+ \cos(2 m_\phi T + 4 \delta) \sin(2 m_\phi T)\tan(m_\phi \Delta t) \biggr\}~.
\ee
Using these expressions it is possible to calculate the $\chi^2$ statistic for any choice of binning (for any value $n_{\rm bin}$).  Given $\Delta t,\, T,\, N_{\rm tot},$ and $\alpha$, we obtain a 90\% confidence sensitivity estimate on $\mathcal{B}_{\rm sig}$ by requiring the branching ratio to be non-negative and utilizing the one-sided upper bound obtained by solving
$\chi^2[\mathcal{B}_{\rm sig}] = Z_{90}$, where $Z_{90} = 2.706$ is the solution to $\textrm{erf}(\sqrt{Z_{90}/2})\equiv 0.9$ -- for more details see Sec.~\ref{sec:time-indep:signal}.

The behavior of numerical results in Figs.~\ref{fig:upper_limit_vs_delta_t} and \ref{fig:upper_limit_vs_delta_t_BelleII} can be understood by considering various limiting cases: 
\begin{enumerate}
\item\emph{Single bin.}
First, let us consider a single time bin analysis, so that one does not resolve the oscillating signal. In this case, the $\chi^2$ statistic in Eq.~\eqref{eq:oschypo} evaluates to
\be\label{eq:onebinchisq}
\chi^2_1 = \frac{\mathcal{B}^2_{\rm sig} f_{\rm sig}^2 N_{\rm tot}}{\mathcal{B}_{\rm bg} f_{\rm bg}(1+\alpha^2 \mathcal{B}_{\rm bg} f_{\rm bg} N_{\rm tot})}\left(1+\frac{\sin \big(2 m_\phi T+2 \delta)-\sin 2\delta}{2m_\phi T}\right)^2~.
\ee
It is useful to compare $\chi_1^2$ with the $\chi^2$ for a signal constant in time, 
\begin{equation}
\label{eq:chi_sqr_static}
    \chi^2_{\rm const}[\mathcal{B}] = \frac{n_{\rm bin} S_{\rm bin}^2}{\sigma_{\rm stat}^2+n_{\rm bin}\sigma_{\rm sys}^2}=\frac{(\mathcal{B} f_{\rm sig} N_{\rm tot})^2}{N_{\rm tot} \mathcal{B}_{\rm bg} f_{\rm bg}+\alpha^2 (N_{\rm tot} \mathcal{B}_{\rm bg} f_{\rm bg})^2}  
\end{equation}
where $\mathcal{B}$ is the time-averaged $\ell_i \to \ell_j \phi'$ branching ratio.  The first equality follows from Eq.~\eqref{eq:chi2} for a constant signal in each bin, $S_i =S_{\rm bin}$, while the second equality uses that $S_{\rm bin}= \mathcal{B} f_{\rm sig} N_{\rm tot}/n_{\rm bin}$. In the limit of fast oscillations, i.e., for the case where the duration of the experiment is much longer than the oscillation period, $m_\phi T\gg 1$, one has $\mathcal{B}_{\rm sig}=\left<2 \mathcal{B}_{\rm sig}\cos^2(m_\phi t+\delta)\right>$, giving for $\chi_1^2$ in Eq.~\eqref{eq:onebinchisq}
\begin{equation}
\label{eq:chi2:avg}
    \chi^2_1\rightarrow \chi^2_{\rm const}[\mathcal{B}_{\rm sig}].
\end{equation}
That is, averaging $\chi_1^2$ over time gives the value of $\chi^2$ for a time-independent signal. 
In the limit of slow oscillations, $m_\phi T\ll 1$, Eq.~\eqref{eq:onebinchisq} becomes
\begin{equation}
\begin{split}
    \chi^2_1~\rightarrow ~\chi^2_{\rm const}[\mathcal{B}_{\rm sig} \cos^2(\delta)] + \mathcal{O}(m_\phi T).
\end{split}
\end{equation}
The $\chi^2$ test statistic is then sensitive to the phase $\delta$ of the classical field $\phi_c$, resulting in a $\chi^2$ that may be larger or smaller than the time-averaged signal with amplitude $\mathcal{B}_{\rm sig}$. For $\delta = \pi/2$ the signal (and thus the $\chi^2$) vanishes at leading order in $m_\phi T$; the first non-zero contribution to the signal is of order $(m_\phi T)^4$, resulting in a highly suppressed, but non-zero $\chi^2$.

\item\emph{Coarse binning.}
If the signal oscillates in the duration of the experiment ($m_\phi T\gg 1$), but the oscillation time is shorter than the bin size ($m_\phi \Delta t \gg 1$) the sensitivity to oscillations is decreased.  In this limit the result for $\chi^2$ is the same as the single bin result, Eq.~\eqref{eq:chi2:avg}, because the binning is too coarse 
\be\label{eq:coursebinning}
\chi^2_{\rm cb} =\chi^2_{\rm const}[\mathcal{B}_{\rm sig}].
\ee
\item\emph{Fine binning.}
If the binning is sufficiently small to resolve the oscillations ($m_\phi \Delta t\ll 1$) and the experiment runs long enough ($m_\phi T\gg 1$), then the $\chi^2$ becomes
\be\label{eq:finebinning}
\chi^2_{\rm fb} = \frac{3}{2} \left(1+\frac{\alpha^2 N_{\rm tot} \mathcal{B}_{\rm bg} f_{\rm bg}}{3}\right)\chi^2_{\rm const}[\mathcal{B}_{\rm sig}]= \frac{3}{2} \left(1+\frac{n_{\rm bin}\sigma_{\rm sys}^2}{3\sigma_{\rm stat}^2}\right)\chi^2_{\rm const}[\mathcal{B}_{\rm sig}].
\ee
When the systematic uncertainties can be ignored ($\alpha^2 N_{\rm tot}f_{\rm bg}\ll 1$) this fine binning performs slightly better than the course binning, $\chi^2_{\rm fb}= 3\chi^2_{\rm cb}/2$.  
When systematic uncertainties dominate, however, fine binning does substantially better, resulting in a  $\chi^2$ that is not limited by the systematic uncertainty, 
\be\label{eq:finebinhighsys}
\chi^2_{\rm fb} = \frac{1}{2N_{\rm tot} \mathcal{B}_{\rm bg} f_{\rm bg}}\left(\mathcal{B}_{\rm sig} f_{\rm sig} N_{\rm tot}\right)^2=  \frac{n_{\rm bin}}{2} \frac{S_{{\rm avg},1}^2}{\sigma_{\rm stat}^2},
\ee
where $S_{{\rm avg},1} \equiv \left<S_i\right> =\mathcal{B}_{\rm sig} f_{\rm sig} N_{\rm tot}/n_{\rm bin}$.
This result is intuitive: it is possible to measure a time-dependent signal even in the presence of large systematic uncertainties, as long as all the systematics are time independent. 
\end{enumerate} 

\subsubsection{ Incoherent (``fast'') oscillations}
\label{sec:multiple:patches}

For a given DM mass $ m_\phi $, the galactic halo’s typical velocity dispersion sets a characteristic coherence time, $\tau_{\rm coh} \simeq (m_\phi v^2)^{-1}$, during which the classical DM background remains effectively in phase. When 
$m_\phi \gtrsim (v^2 T)^{-1}$, the total observation time $T$ spans multiple coherence patches. To estimate the sensitivity to a DM signal in this regime, we account for decoherence by subdividing the observation period into $n_{\rm patch} = \lceil T/\tau_{\rm coh} \rceil = \lceil T\, m_\phi\, v^2 \rceil \simeq \lceil 10^{-6}\, m_\phi T \rceil$ equally spaced patches, where $\lceil \cdots \rceil$ represents the ceiling function. 

{Each patch $ p \in \{1, 2, \dots, n_{\rm patch}\} $ has an unknown phase $ \delta_p $ that must be treated as a nuisance parameter. Within each patch, the phase is approximately constant due to the coherence of the DM field, but the phases in different patches are statistically independent. This independence implies that the tests performed in each patch are themselves independent statistical tests of the signal hypothesis.
To properly account for the multi-patch structure, we compute the test statistic in each patch after setting the phase to its worst-case (most conservative) value $\delta_p^{\rm cons}$, the value that minimizes that patch's $\chi^2$ contribution. 
In the case $T / \tau_{\rm coh} \in \mathbb{Z}$, the $\chi^2$ is minimized when each patch maintains the same phase $\delta_{\rm patch}$, corresponding to the minimization with respect to $\delta$ over a single patch. 
For simplicity, in each patch we set $\delta_p = \delta_{\rm patch}$ despite the possibility of a single partial patch not fully residing within the observation window $T$ (assuming that the first patch starts at the beginning of the observation time). Including this additional minimization on the final partial patch does not meaningfully impact our derived sensitivities.
Under the null hypothesis, the test statistic from each patch is $\chi^2$-distributed with one degree of freedom. 
By Fisher's theorem for combining independent tests, summing these independent $\chi^2_1$ contributions yields a combined statistic
\begin{equation}
\label{eq:chi2_npatch}
\chi^2_{n_{\rm patch}} = \sum_{p=1}^{n_{\rm patch}} \chi^2_p[\delta_p^{\rm cons}],
\end{equation}
which follows a $\chi^2$ distribution with $n_{\rm patch}$ degrees of freedom.}
 
To further simplify the analysis, we restrict the maximum size of each time bin to be the coherence time, $\Delta t \leq \tau_{\rm coh}$, and limit the binning scheme to evenly divide the duration of a single patch, ensuring that no time bin traverses multiple patches. 
Under these approximations, the $\chi^2$ statistic becomes
\begin{equation}\label{eq:multi_patch_chi2}
\begin{split}
  \chi^2 &= \frac{1}{N_{{\rm bg},1}}\sum_{p=1}^{n_{\rm patch}}\sum_{k = 1}^{n^p_{\rm bin}} S_k^2[\delta_p] - \frac{\alpha^2}{1 + \alpha^2 n_{\rm bin} N_{{\rm bg},1}}\left( \sum_{p=1}^{n_{\rm patch}} \sum_{k = 1}^{n^p_{\rm bin}} S_k[\delta_p] \right)^2  \\
  &\simeq n_{\rm patch} \left[ \frac{1}{N_{{\rm bg},1}}\sum_{k = 1}^{n_{\rm bin}^{\rm patch}} S_k^2[\delta_{\rm patch}] - n_{\rm patch} \frac{\alpha^2}{1 + \alpha^2 n_{\rm bin} N_{{\rm bg},1}}\left( \sum_{k = 1}^{n_{\rm bin}^{\rm patch}} S_k[\delta_{\rm patch}] \right)^2 \right] + \chi^2_{\rm partial},
\end{split}
\end{equation}
where the number of bins per patch is given by $n_{\rm bin}^{\rm patch} = \tau_{\rm coh} / \Delta t = T / n_{\rm patch} \Delta t = (m_{\phi} v^2 \Delta t)^{-1}$ and $\chi^2_{\rm partial}$ refers to the additional contribution to the $\chi^2$ stemming from partial patches at the end of the observation time. Because of the fixed binning within each patch, a given $\Delta t$ is not guaranteed to fit within a partial patch, introducing an error from integrating the signal over an additional amount of time $dt_{\rm extra}$ equivalent to
\beq
    dt_{\rm extra} = \left(\left\lceil \dfrac{\tau_{\rm coh}(1 - n_{\rm patch}) + T}{\Delta t} \right\rceil - \dfrac{\tau_{\rm coh}(1 - n_{\rm patch}) + T}{\Delta t}\right) \Delta t~,
\eeq
which vanishes as $\Delta t \to 0$ ($n_{\rm bin}^{\rm patch} \to \infty$). In the numerical examples presented in the main text, this error is accounted for explicitly in the computation of $\chi^2_{\rm partial}$ via subtraction of the extra contribution introduced by $dt_{\rm extra}$. 
{To set 90\% confidence limits in the multi-patch regime, we require $\chi^2_{n_{\rm patch}} \leq Z_{90}(n_{\rm patch})$, where $Z_{90}(n_{\rm patch})$ is the 90th percentile of a $\chi^2$ distribution with $n_{\rm patch}$ degrees of freedom. For large $n_{\rm patch}$, this threshold is well-approximated by
\beq
\label{eq:Z90_npatch}
Z_{90}(n_{\rm patch}) \simeq n_{\rm patch} + 1.28\sqrt{2 n_{\rm patch}},
\eeq
which follows from the normal approximation to the $\chi^2$ distribution.}

\subsubsection{Time-independent signal} 
\label{sec:time-indep:signal}
The $\chi^2$ square statistic for a time-independent $\ell_i \to \ell_j \phi'$ branching ratio $\mathcal{B}_{\rm const}$ is given by $\chi^2_{\rm const}[\mathcal{B}_{\rm const}]$ in Eq.~\eqref{eq:chi_sqr_static}.
Again requiring $\mathcal{B}_{\rm const}$ to be non-negative the one-sided upper bound at $X\%$ confidence level (CL) for $\mathcal{B}_{\rm const}$, denoted as $\mathcal{B}_{\rm const}^{X\% \text{CL}}$, is obtained by solving 
\beq
\chi^2_{\rm const}[\mathcal{B}_{\rm const}] = Z_X \quad \textrm{with}\quad\textrm{erf}(\sqrt{Z_{X}/2})\equiv (X/100).
\eeq
For instance, $Z_{90}=2.706$ and $Z_{95}=3.841$.
This yields
\beq
\label{eq:B_static_CL}
  \mathcal{B}_{\rm const}^{X\% \text{CL}} ~=~ \frac{1}{f_{\rm sig}}\sqrt{Z_{X} \mathcal{B}_{\rm bg} f_{\rm bg} \left( \frac{1}{N_{\rm tot}} + \mathcal{B}_{\rm bg} f_{\rm bg} \alpha^2 \right)}~=~\frac{1}{f_{\rm sig} N_{\rm tot}} \sqrt{Z_Xn_{\rm bin}(\sigma_{\rm stat}^2+n_{\rm bin}\sigma_{\rm sys}^2)}.
\eeq
In the main text, Eq.~\eqref{eq:B_static_CL} was used to determine the systematic uncertainty parameter $\alpha$ for different experiments, given a quoted upper bound $\mathcal{B}_{\rm const}^{X\% \text{CL}}$. For instance, for a systematic-dominated analysis ($\alpha^2 N_{\rm tot} \mathcal{B}_{\rm bg} f_{\rm bg}\gg 1$), Eq.~\eqref{eq:B_static_CL} gives $ \mathcal{B}_{\rm const}^{X\% \text{CL}}  \simeq \alpha \sqrt{Z_X}(\mathcal{B}_{\rm bg} f_{\rm bg} / f_{\rm sig})$.

\subsubsection{Discussion of time-dependent results} 
\label{sec:result}

For time-dependent analysis we show in Fig.~\ref{fig:upper_limit_vs_delta_t} in the main text, and in Fig. \ref{fig:upper_limit_vs_delta_t_BelleII} in the Supplemental Material,  the derived upper limits on the signal branching ratio of $\mu\to e\phi', \tau \to \mu \phi'$ and $\tau \to \ell \phi'$, respectively, as a function of the bin size in time $\Delta t$ for different DM masses. For each mass and $\Delta t$, the phase $\delta$ is chosen by profiling the $\chi^2$ over $\delta$ and selecting the value that minimizes the test statistic, ensuring the most conservative estimate. 
We see that for coarse time bins, in which the oscillatory nature of the signal cannot be resolved ($m_\phi \Delta t \gg 1$), the derived sensitivity is systematics limited. As $\Delta t$ decreases and the time bins become finer, the oscillations start to be resolved, eventually becoming fully resolved around $\Delta t \sim m_\phi^{-1}$ (indicated by the vertical dotted lines). In this regime ($m_\phi \Delta t \ll 1$), the sensitivity is no longer systematics limited but instead becomes constrained by statistics. 
 
The translated sensitivities on the scale $f$ as a function of DM mass, using the most optimal time binning for each mass within the region of validity (between the gray regions denoted in Fig.~\ref{fig:upper_limit_vs_delta_t}), can be found in Fig.~\ref{fig:test}. That is, in the limit of fine-binning, given $T, N_{\rm tot}, \mathcal{B}_{\rm bg}, f_{\rm bg}$, $f_{\rm sig}$, and $\alpha$, the 90\% CL sensitivity estimates on $\mathcal{B}_{\rm sig}$ are obtained by demanding that $\chi^2 \leq Z_{90}$. Using  Eqs.~\eqref{eq:li:to:lj:phi} and \eqref{eq:finebinning}, this requirement translates in the lower bound 
\beq 
\begin{split}
  \frac{f}{\sqrt{C_{ij}}} \geq & \frac{8.9 \times 10^{-12} \,\text{GeV}}{\sqrt{m_\phi}}  \left( \frac{m_{\ell_i}^3}{\Gamma_{\ell_i}} \right)^{1/4} \left(\frac{N_{\rm tot} f_{\rm sig}^2}{\mathcal{B}_{\rm bg} f_{\rm bg}} \left[ \frac{3 + \mathcal{B}_{\rm bg} f_{\rm bg} \alpha^2 N_{\rm tot}}{1 + \mathcal{B}_{\rm bg} f_{\rm bg} \alpha^2 N_{\rm tot}} \right]\right)^{1/8},
\end{split}
\eeq
where $\Gamma_{\ell_i}$ denotes the SM decay width of $\ell_i$. Likewise in the limit of coarse-binning (or a single-bin analysis) the lower limit on $f$ is given by 
\beq 
\begin{split}
  \frac{f}{\sqrt{C_{ij}}} \geq & \frac{9.7 \times 10^{-12} \,\text{GeV}}{\sqrt{m_\phi}}  \left( \frac{m_{\ell_i}^3}{\Gamma_{\ell_i}} \right)^{1/4} \left(\frac{N_{\rm tot} f_{\rm sig}^2}{\mathcal{B}_{\rm bg} f_{\rm bg}(1 + \mathcal{B}_{\rm bg} f_{\rm bg} \alpha^2 N_{\rm tot})}\right)^{1/8}.
\end{split}
\eeq

Note that the derivation leading to Eq.~\eqref{eq:finebinning} has several implicit assumptions, all of which can fail:
\begin{itemize}
  \item We assume that the data is collected continuously over the full observation time. More realistic experimental setups, where the observation window is broken into discrete chunks of continuous data-taking, can be taken into account with minimal and straightforward modifications to the analysis described above. Alternatively one could also use the Lomb-Scargle periodogram to search for the time-dependent signal \cite{VanderPlas_2018}. 
  \item {So far, in the $\chi^2$ approach, we assumed that each time bin contains a sufficient number of events such that the statistic is $\chi^2$ distributed, i.e., that each bin contains at least 10 events (background + signal). 
 The requirement of $10$ events per bin is a simplification to allow the straightforward application of $\chi^2$ statistics.  Smaller periods (larger masses) can be considered at the expense of using slightly more advanced statistical techniques such as unbinned log-likelihood based analyses or Rayleigh periodograms~\cite{Losada:2023zap}.  We describe the Rayleigh periodogram approach in Sec.~\ref{sec:Rayleigh}.  In Figs.~\ref{fig:upper_limit_vs_delta_t} and \ref{fig:upper_limit_vs_delta_t_BelleII} we use the the $\chi^2$ approach for low masses (large $\Delta t$) and transition to the Rayleigh approach when the number of events per bin drop below 10.   There is an intrinsic limit on the frequency an experiment is sensitive to, which is determined by the resolution of their event timing.  }
  \item In Eq.~\eqref{eq:finebinning} we assume that the DM phase is fully coherent over the full observation time $T$. In a full experimental analysis, one must also properly account for the effect of phase coherence.
  As described in Sec.~\ref{sec:multiple:patches}, in our analysis, we account for DM decoherence approximately by considering $n_{\rm patch} = T / \tau_{\rm coh}$ equally spaced coherent patches where each patch maintains the same phase over its duration, chosen conservatively through a minimization of the test statistic over $\delta$. 
\end{itemize}

\subsubsection{{Rayleigh Periodogram}}
\label{sec:Rayleigh}

The Rayleigh periodogram avoids the need to keep time bins large enough to contain $\sim 10$ 
events, and thus allows sensitivity to higher dark matter masses than the $\chi^2$ approach.  It can be thought of as an unbinned analysis or, equivalently, as an analysis where the time bins are the size of the experiment's timing resolution $\tau_{\rm res}$ and each such bin contains at most 1 event.  

The Rayleigh power spectrum, at frequency $f$, is defined \cite{Losada:2023zap} as 
\begin{equation}
z\left(f\right)=\frac{2}{N_e}\left(\left[\sum_{n=1}^{N_e}\cos\left(2\pi f t_{n}\right)\right]^{2}+\left[\sum_{n=1}^{N_e}\sin\left(2\pi f t_{n}\right)\right]^{2}\right)~,
\label{eq:appRayleigh}
\end{equation}
where the times $t_n$ are the times of the $N_e$ events.  Within a time bin of size $\Delta t = \tau_{\rm res}$ the expected number of events,  $\dot{N}_{\rm pred}(t) \tau_{\rm res}$, is small.  The probability of seeing $1$ event in the bin, assuming dark matter modulates decays, is
\be
P(1\ \mathrm{event}) =  \frac{N_{\rm tot} \tau_{\rm res}}{T} \big[\mathcal{B}_{\rm bg} f_{\rm bg} 
  + 2 \mathcal{B}_{\rm sig} f_{\rm sig} \cos^2 (m_\phi t + \delta) \big]~.
\ee
From this one can calculate the expected Rayleigh power at $f=2m_\phi$.  If the coherence time is longer than the duration of the experiment and there are many oscillations during the experiment, i.e. $\tau_{\rm res} \gg T \gg 1/m_\phi$, then the expected Rayleigh power is insensitive to the phase of the dark matter: 
\be\label{eq:RPsinglephase}
\langle z \rangle = 2+\frac{\mathcal{B}^2_{\rm sig} f^2_{\rm sig}}{\mathcal{B}_{\rm bg} f_{\rm bg} + \mathcal{B}_{\rm sig} f_{\rm sig}}\frac{N_{\rm tot} }{2}~.
\ee

Finally, under the background only hypothesis the Rayleigh power is distributed as a $\chi^2$ with 2 degrees of freedom, which means that 
\begin{equation}
P(z>Z) = e^{-Z/2}~,
\end{equation}
and the $90~\%$ C.L.~corresponds to $Z^{\mathrm R}_{90}=4.605$.  Comparing (\ref{eq:RPsinglephase}) to (\ref{eq:finebinhighsys}) we see that the 90\%C.L.~on the time-dependent branching ratio is very slightly different using the $\chi^2$ and Rayleigh power approaches.  This slight difference leads to a small discontinuity in the bounds as we transition from one approach to the other as the mass grows.

If the coherence time is shorter than the length of the experiment there is a suppression in the Rayleigh power due to the phase of the dark matter changing over a coherence time.  This suppression is encoded by the replacement $N_{\rm tot}\rightarrow  N_{\rm tot}/n_{\rm patch}$ in (\ref{eq:RPsinglephase}).
It is instructive to compare how the sensitivity of both the $\chi^2$ and Rayleigh approaches scale in the incoherent regime where $T \gg \tau_{\rm coh}$. For the $\chi^2$ in the fine-binned, systematics-dominated limit, the 90\% CL threshold grows as $Z_{90}(n_{\rm patch}) \simeq n_{\rm patch}$ for $n_{\rm patch} \gg 1$, yielding a sensitivity
\beq
\mathcal{B}_{\rm sig}^{90\%,\chi^2} = \sqrt{\frac{2 Z_{90}(n_{\rm patch}) \mathcal{B}_{\rm bg} f_{\rm bg}}{f_{\rm sig}^2 N_{\rm tot}}} \propto n_{\rm patch}^{1/2}.
\eeq
For the Rayleigh periodogram, the expected power receives a suppression $N_{\rm tot} \to N_{\rm tot}/n_{\rm patch}$ in Eq.~(\ref{eq:RPsinglephase}), giving
\beq
\mathcal{B}_{\rm sig}^{90\%,\text{R}} = \sqrt{\frac{2(Z_{90}^{\text{R}} - 2) n_{\rm patch} \mathcal{B}_{\rm bg} f_{\rm bg}}{f_{\rm sig}^2 N_{\rm tot}}} \propto n_{\rm patch}^{1/2}.
\eeq
Both test statistics exhibit the same $n_{\rm patch}^{1/2}$ degradation in sensitivity.

This scaling has an important consequence: for sufficiently fast oscillations (large $m_\phi$), the time-dependent analysis loses its advantage over the time-independent single-bin approach. The latter, in the systematics-dominated regime, has sensitivity $\mathcal{B}_{\rm const}^{90\%} \simeq \alpha \sqrt{Z_{90}} (\mathcal{B}_{\rm bg} f_{\rm bg}/f_{\rm sig})$ from Eq.~\eqref{eq:B_static_CL}, which is independent of $n_{\rm patch}$. Once the number of coherence patches becomes large, the growing $n_{\rm patch}^{1/2}$ penalty erodes the benefit of time-binning, and the time-independent limit ultimately provides comparable sensitivity. This behavior is reflected in the flattening of the sensitivity curves at larger values of $m_\phi$ in Fig.~\ref{fig:test}.

\subsubsection{Experimental details} 
In this subsection we give further details about the treatment of experimental projections. 

\begin{itemize}
\item
\textbf{Mu3e $\boldsymbol\mu \boldsymbol \to \boldsymbol e  \boldsymbol \phi'$}:  A defining feature of the $\mu\to e\phi'$ signal are monoenergetic positrons at energy $E_e\simeq m_\mu/2$.  This mono-energetic line lies at the endpoint of the SM decay $\mu\to e\nu\bar{\nu}$ ($\mathcal{B}_{\mu \to e \nu \bar{\nu}} = \mathcal{B}_{\rm bg} \simeq1$), given by the kinematic configuration of two neutrinos collinear that are back-to-back with the positron. The size of the background from the SM is determined by the detector's energy resolution. In our projections we take the positrons to be in the endpoint region, if their energy is within $3~\rm MeV$ of the kinematic endpoint; this choice corresponds to the anticipated detector's intrinsic energy resolution~\cite{Mu3e:2020gyw,Perrevoort:2018okj, Perrevoort:2023qhn}, and is also comparable with the signal region used in the analysis in \cite{Banerjee:2022nbr}, where it was defined to be within $\simeq 4$\,MeV of the endpoint, but also with a cut on the direction of positrons relative to muon spin. In this narrow window, the estimated background level is $N_{\rm bg} \simeq 1.1 \times 10^{13}$, which is what we use in our projections. For comparison, widening the endpoint energy window to twice the resolution (i.e.,~$6\,\rm MeV$) would increase the background yield to approximately $5.7\times10^{13}$ events. Assuming, for simplicity, that the detection and reconstruction efficiencies are $\simeq1$, we have $f_{\rm bg}\simeq N_{\rm bg} / N_{\rm tot} = 3.3\times 10^{-3}$. 
We also assume that the systematic uncertainties are dominated by the theory prediction on the SM background. This gives an expected 90\%CL upper limit on $\mu\to e\phi'$ branching ratio of $\mathcal{B}^{\rm 90\%}_{\rm const}= 6\times10^{-7}$ (as determined in Ref.~\cite[Fig.~12]{Banerjee:2022nbr}).
From it we deduce the systematic uncertainty parameter $\alpha$, assuming that approximately all of the signal events fall into this kinematic endpoint ($f_{\rm sig} \simeq 1$) as well as systematics domination ($\alpha^2 N_{\rm tot} \mathcal{B}_{\rm bg} f_{\rm bg}\gg 1$) in Eq.~\eqref{eq:B_static_CL}, giving $\alpha \simeq (\mathcal{B}^{\rm 90\%}_{\rm const} f_{\rm sig})/(\sqrt{Z_{90}}\,\mathcal{B}_{\rm bg}f_{\rm bg})=1.1 \times 10^{-4}$.

\item \textbf{Belle-II} $\boldsymbol \tau \boldsymbol\to \boldsymbol\ell \boldsymbol\phi'$: 
During the 2019-2020 run, Belle-II recorded roughly $N_{\rm tot} \simeq 1.2\times10^8$ taus~\cite{Belle-II:2022heu} (we assume that these were collected continously over 300\,days). 
Of these, approximately ${\cal B}_{\tau \to e \nu \bar{\nu}}=17.8\%$ (for $\ell=e$) and $\mathcal{B}_{\tau \to \mu \nu \bar{\nu}} = 17.4\%$ (for $\ell=\mu$) decay via the SM channel $\tau\to\ell\nu\overline{\nu}$ \cite{ParticleDataGroup:2024cfk}.
The irreducible SM background thus has a branching of 
\beq
\label{eq:Bbg}
\mathcal{B}_{\rm bg}^{e[\mu]} \simeq 0.178\,[0.174] \left( \frac{\mathcal{B}_{\tau \to \ell \nu \bar{\nu}}}{0.178\,[0.174]} \right).
\eeq
In the tau pseudo-rest frame the distribution of charged lepton energy $E_\ell^*$ for these SM decays resembles the one expected from the exotic $\tau\to\ell\phi'$ decays, since the poor reconstruction of the tau rest frame smears the mono-energetic signature. In Belle-II shape information was used to distinguish between background and signal. In our estimates we instead use a simplified strategy, and define signal regions as the energy windows $E_e^{*}/(m_\tau/2)\in[0.66,1.48]$ and $E_\mu^{*}/(m_\tau/2)\in[0.71,1.38]$ defined to retain $\epsilon_{\rm cut}^{\rm sig} \simeq 90\%$ of the $\tau\to\ell\phi'$ signal events, $N_{\rm sig}$, and straddle symmetrically the peak of the signal distribution. These windows simultaneously capture about $\epsilon_{\rm cut}^{\mathrm{bg}, e} \simeq 59\%$ and $\epsilon_{\rm cut}^{\mathrm{bg}, \mu} \simeq 56\%$ of the SM background events, $N_{\rm bg}$, for electrons and muons, respectively.
Moreover, the minimal reconstruction efficiency for the signal channel is estimated at $\epsilon_{\rm rec}^{{\rm sig}, e} \simeq 13.4\%$ for electrons and $\epsilon_{\rm rec}^{{\rm sig}, \mu} \simeq 17.4\%$ for muons~\cite[Table 4.4]{Hernandez:2023foy}, while for the background $\epsilon_{\rm rec}^{{\rm bg}, e} \simeq 12.7\%$ and $\epsilon_{\rm rec}^{{\rm bg}, \mu} \simeq 16.2\%$ \cite{Belle-II:2022heu,Hernandez:2023foy}. Finally, the Belle-II analysis also requires the tag hemisphere to contain three charged particles, $\tau \to 3h \nu_\tau$, $h = \pi, K$, with a branching $\mathcal{B}_{\tau \to 3h\nu} = 15.2\%$ \cite{ParticleDataGroup:2024cfk}. Because both the signal and background processes will be subject to this requirement, we incorporate this as a selection efficiency, $\epsilon_{\rm sel} \equiv \mathcal{B}_{\tau \to 3h\nu}$.
Combining the kinematic cuts and the reconstruction efficiencies yields 
\begin{align}
    f_{\rm bg}^{e[\mu]} &\simeq 1.1\times 10^{-2}\, [1.4\times 10^{-2}]\left( \frac{\epsilon_{\rm cut}}{0.59 [0.56]}\right) \left( \frac{\epsilon_{\rm rec}}{0.127 [0.162]} \right) \left( \frac{\epsilon_{\rm sel}}{0.152} \right),
\\
    f_{\rm sig}^{e[\mu]} &\simeq 1.8\times 10^{-2}\, [2.4\times 10^{-2}] \left( \frac{\epsilon_{\rm cut}}{0.9}\right) \left( \frac{\epsilon_{\rm rec}}{0.134 [0.174]} \right) \left( \frac{\epsilon_{\rm sel}}{0.152} \right).
\end{align}
These agree within 30\% with the values one would  obtain if the results in Fig.~1 in \cite{Belle-II:2022heu} were used instead.
Using the relation $\alpha\simeq (\mathcal{B}^{\rm 90\%}_{\rm const} f_{\rm sig})/(\sqrt{Z_{90}}\,\mathcal{B}_{\rm bg} f_{\rm bg})$ in Eq.~\eqref{eq:B_static_CL}
with the 90$\%$ confidence-level limits $\mathcal{B}^{\rm 90\%}_{\rm const}\simeq 7.6\times10^{-4}$ for electrons and $4.7\times10^{-4}$ for muons~\cite[Tab.~III]{Belle-II:2022heu}, we obtain systematic uncertainty parameters of $\alpha \simeq 2.7 \times 10^{-2}$ and $\alpha \simeq 1.9 \times 10^{-2}$ for the electron and muon channels, respectively.
For the full Belle-II analysis utilizing the entire $50\,\text{ab}^{-1}$ integrated luminosity dataset, assumed to be taken over $T\approx 3000$ days continuously, we provide sensitivity projections using the same analysis outlined above (with an appropriately rescaled $N_{\rm tot}$) assuming that the time integrated analysis will remain systematics dominated at the present value. While this is very likely to be too conservative, given that improvements in both the systematics and the observables used to distinguish between signal and background are likely \cite{DeLaCruz-Burelo:2020ozf}, the approximation suffices for our purposes: the rough estimate of the reach for the time-dependent analysis. 

\item 
\textbf{FCC-ee} $\boldsymbol \tau \boldsymbol\to \boldsymbol\ell \boldsymbol\phi'$ : While running in Tera-Z mode the proposed FCC-ee experiment will produce approximately $N_{\rm tot} = 2 N_{\tau \bar{\tau}} \simeq 2 \times 1.7\times 10^{11}$ taus over $T=740$ days of running \cite{Bernardi:2022hny}. 
While the changes in the boost of the taus, coverage of the detector, and energy resolution will doubtless lead to changes in signal and background separation, to date there have not been detailed studies of the expected reach for the $\tau \to \ell\phi'$ channel at FCC.  Here, for simplicity, we assume a similar analysis and background contamination to that of Belle-II, and take the reconstruction efficiency to be $\epsilon_{\rm rec}^{\rm sig,bg} \simeq 90\%$.  Ignoring small differences between $e$ and $\mu$ this gives
\beq
    f_{\rm bg} \simeq 8.1\times10^{-2}\left( \frac{\epsilon_{\rm cut}}{0.59}\right) \left( \frac{\epsilon_{\rm rec}}{0.9} \right) \left( \frac{\epsilon_{\rm sel}}{0.152} \right),
\eeq \vspace{-0.2in}
\beq
    f_{\rm sig} \simeq 0.12 \left( \frac{\epsilon_{\rm cut}}{0.9}\right) \left( \frac{\epsilon_{\rm rec}}{0.9} \right) \left( \frac{\epsilon_{\rm sel}}{0.152} \right).
\eeq
As a surrogate for the expected bound on $\tau\to \ell \phi'$ we use the expected systematics dominated (absolute) precision on the leptonic branching fraction of $\sigma (\mathcal{B}_\ell)=3\times 10^{-5}$.   It is hoped that, even if the final precision is systematically limited, it will not be far from this  value \cite{Dam:2021ibi}.  Using $\alpha \simeq  \sigma (\mathcal{B}_\ell) f_{\rm sig}/(\mathcal{B}_{\rm bg}f_{\rm bg})$ with $\mathcal{B}_{\rm bg}$ in Eq.~\eqref{eq:Bbg}, we find $\alpha\simeq 2.6 \times 10^{-4}$.
\end{itemize}

\end{document}